\documentclass[
 reprint,
 amsmath,
 pre,
 amssymb,
 nofootinbib,
 aps,
]{revtex4-2}

\usepackage[pdftex]{graphicx}
\usepackage{wasysym}
\usepackage{amsfonts}
\usepackage{ifsym}

\usepackage{amsmath,amssymb,amsfonts,amscd,mathrsfs}
\usepackage{dsfont}

\usepackage{lipsum}
\usepackage{pifont}

\usepackage{tikz-feynman}
 \tikzfeynmanset{compat=1.0.0}

\makeatletter
\newlength{\apb@width}
\newcommand{\autoparbox}[2][c]{\settowidth{\apb@width}{#2}\parbox[#1]{\apb@width}{#2}}

\makeatother

\newcommand{\namedref}[2]{\hyperref[#2]{#1~\ref*{#2}}}

\newcommand{\be}{\begin{equation}}
\newcommand{\ee}{\end{equation}}

\newcommand{\reef}[1]{(\ref{#1})}

\def\bea#1\eea{\begin{align}#1\end{align}}

\newcommand{\diagUVtwo}{
  \begin{minipage}[h]{0.09\linewidth}\begin{tikzpicture}
  [
roundnode/.style={circle, draw=black!60, fill=black!6, very thick, 
  inner sep=2.1pt,
  text width=3mm},
roundnode3/.style={circle, draw=black!60, fill=black!6, very thick, 
  inner sep=2.5pt,
  text width=3mm},
]
\begin{feynman}[small]
\node [dot] (X) at (-.4,0);
\vertex (x1l) at (-.8,.3);
\vertex (xnl) at (-.8,-.3);
\node [dot] (Y) at (0,0);
\vertex (X1l) at (.4,.3);
\vertex (Xnl) at (.4,-.3);
   \diagram*{
   (x1l) -- [scalar, thick, quarter left, looseness=.5] (X) --[ scalar, thick, quarter left, looseness=.5] (xnl) ,
      (Y) -- [very thick] (X) ,
     (X1l)   -- [scalar, thick, quarter right, looseness=.5] (Y)  -- [scalar, thick, quarter right, looseness=.5] (Xnl)  ,
};
 \end{feynman}
\end{tikzpicture}
  \end{minipage} 
  }

\newcommand{\diagIRone}{
  \begin{minipage}[h]{0.09\linewidth}\begin{tikzpicture}
  [
roundnode/.style={circle, draw=black!60, fill=black!6, very thick, 
  inner sep=2.1pt,
  text width=3mm},
roundnode3/.style={circle, draw=black!60, fill=black!6, very thick, 
  inner sep=2.5pt,
  text width=3mm},
]
\begin{feynman}[small]
\node [dot] (X) at (-.4,0);
\vertex (x1l) at (-.8,.3);
\vertex (xnl) at (-.8,-.3);
\node [dot] (Y) at (.2,0);
\vertex (X1l) at (.6,.3);
\vertex (Xnl) at (.6,-.3);
   \diagram*{
   (x1l) -- [scalar, thick, quarter left, looseness=.5] (X) --[ scalar, thick, quarter left, looseness=.5] (xnl) ,
      (Y) -- [scalar, thick, half left, looseness=1.2] (X) -- [scalar, thick,  half left, looseness=1.2] (Y) ,
     (X1l)   -- [scalar, thick, quarter right, looseness=.5] (Y)  -- [scalar, thick, quarter right, looseness=.5] (Xnl)  ,
};
 \end{feynman}
\end{tikzpicture}
  \end{minipage} 
  }

\newcommand{\eps}{\varepsilon}

\renewcommand{\Re}{\mathop{\mathrm{Re}}}
\renewcommand{\Im}{\mathop{\mathrm{Im}}}

\newcommand{\Csphere}{{}^\bullet\kern-1.2pt C}
\newcommand{\Ctorus}{{}^\circ\kern-1.2pt C}

\newcommand{\COMMENT}[1]{}

\newcommand{\neqa}{\nonumber\end{eqnarray}}

\newcommand{\<}{{\langle}}
\renewcommand{\>}{{\rangle}}

\newcommand{\re}{\relax{\rm I\kern-.18em R}}

\def\su2{{SU(2)}}

\def\eps{{\epsilon}}

\def\[{\left[}
\def\]{\right]}

\def\om{\omega}

\def\({\left(}
\def\){\right)}
\def\[{\left[}
\def\]{\right]}

\def\<{\langle}
\def\>{\rangle}

\def\i2{\frac{i}{2}}

\def\2F1{\,_2{\rm F}_1}

\def\black{\color{black}}

\usepackage{color}
\usepackage{graphicx}
\usepackage{dcolumn}
\usepackage{bm}
\usepackage{hyperref}

\usepackage{cancel}
\usepackage{ctable}
\usepackage{booktabs}
\newcolumntype{L}[1]{>{\raggedright\let\newline\\\arraybackslash\hspace{0pt}}m{#1}}
\newcolumntype{C}[1]{>{\centering\let\newline\\\arraybackslash\hspace{0pt}}m{#1}}
\newcolumntype{R}[1]{>{\raggedleft\let\newline\\\arraybackslash\hspace{0pt}}m{#1}}

\newcommand{\beq}{\begin{equation}}
\newcommand{\eeq}{\end{equation}}
\newcommand{\beqq}{\begin{equation*}}
\newcommand{\eeqq}{\end{equation*}}
\newcommand\beqa{\begin{eqnarray}}
\newcommand\eeqa{\end{eqnarray}}
\newcommand\beqaa{\begin{eqnarray*}}
\newcommand\eeqaa{\end{eqnarray*}}

\begin{document}


\title{ 
Extremal Higgs couplings
}

\author{Joan Elias Mir\'o}
\affiliation{The Abdus Salam ICTP,    Strada Costiera 11, 34135, Trieste, Italy }
\author{Andrea L. Guerrieri}
\affiliation{
Dipartimento di Fisica e Astronomia, Universita degli Studi di Padova, Italy \\ Istituto Nazionale di Fisica Nucleare, Sezione di Padova, via Marzolo 8, 35131 Padova, Italy \\
 Perimeter Institute for Theoretical Physics, Waterloo, Ontario N2L 2Y5, Canada 
}
\author{Mehmet As{\i}m G\"{u}m\"{u}\c{s}}
\affiliation{
SISSA, Via Bonomea 265, I-34136 Trieste, Italy 
\\ INFN, Sezione di Trieste, Via Valerio 2, 34127 Trieste, Italy}


\begin{abstract}
We critically assess to what extent it makes sense to bound the Wilson coefficients of dimension-six operators. 
In the context of Higgs physics, we establish that a closely related observable, $c_H$, is well-defined and satisfies a two-sided bound. 
$c_H$ is derived from the low momentum expansion of the scattering amplitude, or the derivative of the amplitude  at the origin with respect to the Mandelstam variable $s$, 
expressed as  $M(H_iH_i\rightarrow H_jH_j)=c_H s +O(g_\text{SM}, s^{-2})$ where $g_\text{SM}$  represents all Standard Model couplings.
This observable is \emph{non-dispersive} and, as a result, not sign-definite. 
We also determine the conditions under which the bound on $c_H$ is equivalent to a bound on the dimension-six operator $O_H=\partial| H|^2 \partial |H|^2$.

\end{abstract}

\pacs{Valid PACS appear here}
\maketitle


\section{Introduction and conclusions}

The Higgs particle stands out as one of the most enigmatic particles discovered thus far. 
Examining it from every conceivable perspective is a crucial endeavour.
In this note, we initiate the   theoretical study of the  maximal  Higgs coupling strengths. 
Present-day measurements of Standard Model (SM) Higgs couplings exhibit a  good level of agreement with the SM theory. 
Nevertheless, the possibility of new physics emerging at the few TeVs scale remains a compelling avenue for Beyond the Standard Model Physics (BSM).
Despite the potential need for some degree of fine-tuning, this avenue holds the promise of shed\-ding light on the mechanism that governs Electroweak Symmetry Breaking.

Precise measurements of Higgs couplings are of particular significance  in scenarios where the Higgs is a light composite boson. New physics with strong couplings often involves heavy and broad resonances that may prove challenging to directly observe at the LHC. Nevertheless, these resonances can leave their mark as deviations in the Standard Model Higgs couplings.

The deviations are  largest in UV completions of the SM  featuring strong coupling dynamics. 
In this context, the Strongly Interacting Light Higgs~\cite{Giudice:2007fh} (SILH) Effective Field Theory provides valuable insights.
It offers power counting rules for the Higgs EFT, where the Higgs emerges as a light pseudo-Goldstone boson of a strongly interacting sector. The Higgs becomes  massless in the limit $g_\text{SM}\rightarrow 0$ (where  $g_\text{SM}$ collectively denotes the SM couplings) and acquires a small mass through radiative corrections for $g_\text{SM}\neq 0$.

We will make the following simplifying assumption: we consider the UV  BSM couplings to be significantly larger than the SM ones, allowing us to treat the latter as small perturbations.
We should further assume that the composite sector  enjoys a custodial  global symmetry  $SO(4)\simeq SU(2)_L\otimes SU(2)_R$. 
 Under these assumptions, the Higgs sector of SILH simply reads 
${\cal L}_\text{H}^\text{int} = \frac{g_H}{2f^2}\partial_\mu |H|^2\partial^\mu| H|^2+ {\cal O}(f^{-4})$, where we are neglecting operators of dimension (dim.) eight and higher. 
This effective description  can break down either because of strong coupling dynamics at energies above the  scale $f^2$, or because of   the need to incorporate  new resonances in the  perturbative regime of the effective description.

The  dim.-six  operator $\partial_\mu |H|^2\partial^\mu| H|^2=|H|^2 |\partial H|^2+\text{e.o.m.}$ is interesting because, after accounting for the  Higgs vacuum expectation value, it leads to a wave function re-normalisation of the Higgs which in turn results in a universal shift of all SM Higgs couplings.

We emphasise why our chosen simplification is interesting in the context of bounding Higgs coupling deviations: it retains the key complexities of the real problem and allows gradual step-by-step relaxation of these assumptions, paving the way for a more realistic model.

Causality and unitarity  constraints,  
of the two-to-two scattering matrix element  %
imply sharp bounds on the ${\cal L}_H$ EFT.  
The two-to-two scattering  of two $SO(4)$ vectors   can proceed through three different channels
\be
{\bf  M}{=}M   (\bar s | \bar t, \bar u) \delta_{ab}\delta^{cd}{+}
M   (\bar t | \bar u, \bar s) \delta_{a}^{\, c}\delta_{b}^{\, d}{+}
M   (\bar u | \bar s, \bar t) \delta_{a}^{\, d}\delta_{b}^{\, c}  \label{anp}
\ee
annihilation, reflection and transmission of the  vector 
 indices $\{a,b,c,d\}$.   
We will  often   use  shifted   Mandelstam variables  $(\bar s,\bar t, \bar u) \equiv (s,t,u)-4/3 m^2(1,1,1)$.
Recall that momentum conservation implies $s+t+u=4m^2$.
Crossing symmetry dictates that the physical process \reef{anp} can be describes as the boundary value of a single analytic function with the symmetry $M   (\bar s | \bar t, \bar u)= M   (\bar s | \bar u, \bar t)$.

The  unitary and  crossing-symmetry imply a cut that   extends from $(s,t,u)= 4m^2$ to infinity. Therefore the point $(\bar s,\bar t, \bar u) =0$ is analytic and  the amplitude $M(\bar s | \bar t , \bar u)$ can be characterised by its series around the origin,
\be
 M \, / \,  (4 \pi)^2  = c_\lambda + c_H \bar s + c_2  (\bar t^2 + \bar u ^2) + c_2^\prime \bar s^2 +  {\cal O}(\bar s, \bar t, \bar u)^3   \label{amp}
\ee
where the coefficients $c_\lambda$, $c_H$,  $c_2$, $c_2^\prime$ are real valued, and $(4 \pi)^2$ is a convenient normalization. 
The Wilson coefficient $g_H$ describes  the single dim.-six operator  contribution to  \reef{amp} at tree-level:
$
c_\lambda{=} {\cal O}(\frac{m^2}{f^2})$, $c_H{=} \frac{g_H}{(4\pi^2)f^2}{+}{\cal O}(\frac{m^2}{f^4})$.

The main result of this work  is a bound on the parameter $c_H$ in \reef{amp}.
We will discuss two extreme \emph{single energy scale scenarios},
and argue that interesting physics lies in the interpolation of the two.

In the first scenario, sec.~\ref{islando4},    
we look for the extremal values of $c_H$ by making  no assumption   of weak coupling.
We are lead to the rigorous   bound
\be
-0.46 < c_H  \times  m^2  < 1.07    \label{bound1}
\ee
 The single  scale in the problem is the mass $m^2$, thus we can set units $m^2=1$.
This bound is saturated by amplitudes that are maximally strongly coupled  all the way down to the IR  $s\gtrsim4m^2$.
These amplitudes do not feature   an energy  scale separation between $m^2$ and a putative scale of new physics.
Therefore a simple EFT interpretation of \reef{bound1} in terms of operators 
is hardly possible. Although this is not a useful bound for Higgs physics, it is nevertheless an interesting \emph{proof of principle} for the existence of a  universal bound: 
 any theory with the same symmetries must take values within \reef{bound1}.
 In sec.~\ref{pbr} we discuss  how to isolate weakly coupled amplitudes within the space of non-perturbative $O(4)$ theories.

The second scenario, sec.~\ref{eftb}, is complementary and 
 assumes that physics below a new energy scale $\Lambda^2$ is much weaker than new physics above $\Lambda^2$, and  a large scale separation $\Lambda^2\gg m^2$.
In this limit we  are left with a single scale $\Lambda^2$ and we find 
\be
-0.31 < c_H \times \Lambda^2 < 0.35 \label{bound2}
\ee
The bound  \reef{bound2} is saturated by   amplitudes  that on one hand are  maximally strongly coupled above the cutoff scale $\Lambda$,
but on the other hand are very weakly coupled below. 
This limiting case has been dubbed \emph{UV domi\-nated EFTs}.
In this single scale  problem  we can set units $\Lambda^2{=}1$ and interpret  \reef{bound2}
as a universal bound on the space of UV dominated EFTs. 
We will argue that, under certain specific conditions,  this bound can be interpreted in terms of  ${\cal L}_H$ and identify $\Lambda^2=f^2$ and  
$ c_H \times   f^2 = g_H/ (4\pi)^2 $.
In section \ref{eftb} we also    explain how to smoothly interpolate the bounds of the two limiting scenarios \reef{bound1} and \reef{bound2}, see fig.~\ref{fig:interpolation}. 
In section \ref{IRmodel} we show how to    incorporate  IR EFT corrections in order to obtain a more refined bound.

We end this note with a discussion on the interpretation of the bounds in terms of dim.-six operators, sec.~\ref{dim6}, 
and with a final discussion about future directions, sec.~\ref{so}.

We have included a number of  appendices  with details on the calculations and a  \texttt{Mathematica} notebook to streamline the reproducibility.

\section{Dual Bootstrap for \emph{O(\lowercase{n})} theories}
\label{space}

\subsection{Set up and constraints}

We begin  this section by discussing  the constraints on the amplitude \reef{anp} that we are going to use.
The amplitude  satisfies the  double-subtracted dispersion relation
\bea
{\cal A}^I(s,t)&\equiv M^{I}(s,t)- \mathbb{C}^{I}(s,t)  \label{fixedtoN} \\ 
{-}\int_{4}^\infty   & dz    \big[  \mathbb{K}^{IJ}(z;s,t)  M_z^{J}(z,t) {+}  \mathbb{L}^{IJ}(z; s,t) M_z^J(z,z_0)  \big]   {=}0 \nonumber
\eea
where $\vec{\mathbb{C}}(s,t) =  c_\lambda  (n+2 , 2, 0 )+ c_H ( n-1 , -1 , 1)$ $ \bar s  + c_H ( 0 , 0 , 2) \bar t 
$ and  is decomposed into irrep. channels $ M^{(I)} \equiv (M^{(sing)} , M^{(sym)} , M^{(anti)} )$. 
The Kernels  $\mathbb{K}$ and $\mathbb{L}$ are simple rational functions of its arguments; the  derivation of \reef{fixedtoN}  is  given in appendix~\ref{disprel}. In \reef{fixedtoN} and in the rest of this section we set $m^2=1$.

By taking derivatives of  \reef{fixedtoN} one can express any low energy coefficient $c_i$ of \reef{amp}
in terms of a sum rule involving 
integrals over the amplitude's discontinuity.
If  the definition of $c_i$ involves more than two derivatives of the amplitude  with respect to `$s$' the subtraction terms are not present on the sum rule because $\partial_s^2\vec{\mathbb{C}}(s,t)$ in \reef{fixedtoN} vanishes.  In this case the $c_i$'s may enjoy positivity properties that follow from $\text{Im}M^J\geq0$.
Instead, if the definition of $c_i$ involves less than two derivatives of the amplitude, the subtraction  terms are present and thus  the positivity  
of the sum rule is typically spoiled. This is the case of $c_H$
\be
 c_H \,\frac{\pi}{3}  (s-4) =\text{Re}f^{(3)}_1(s)    \hspace{.006cm} -  \hspace{-.05cm}
  \int_{4}^\infty \hspace{-.2cm} dv  \, k^{(3,J)}_{1,\ell}(s,v)  \text{Im}f^{(J)}_\ell(v) 
\label{chsr}
\ee
where  repeated indices $\ell$ and $J$ are summed over and the Kernel $k_{1,\ell}^{3,J}$ is the partial wave projection of $\mathbb{K}$ and $\mathbb{L}$, its exact form is given in appendix~\ref{disprel}, and $f_j^{(I)}$ are the partial wave projections of the amplitude.

Because of the presence of subtraction terms in the  sum rule \reef{chsr}, the value of  $c_H$ is not sign definite. 
Previous works analysed the sign constraints of $c_H$ by means of unsubtracted dispersion relations~\cite{Low:2009di,Falkowski:2012vh} and positivity constraints 
$\text{Im} f_j^{(I)}>0$.

Positivity  constraints  follow from the unitary inequality
\be
{\cal U}_\ell^{(I)}\equiv   2 \text{Im} f_\ell^{(I)}(s)  -\rho(s)\big|  f_\ell^{(I)}(s)   \big|^2 \geq 0 \label{unit1}
\ee
 where $\rho(s)=\sqrt{(s-4)/4}$.
Unitarity constraints  
bound both the real $\text{Re}f_j^{(I)}$ and imaginary  $\text{Im}f_j^{(I)}$  parts of  the amplitude. 
Therefore,  by using the unitarity  constraints  \reef{unit1}  (instead of positivity constraints only)  
we may hope of being able to bound the minimal and maximal value that $c_H$ in \reef{chsr} can attain.
Establishing the existence of this bound is non-trivial,
 as it involves an infinite sum over partial waves on the right-hand side.
 Nevertheless, we will demonstrate in the next section that this hope is indeed realised.

Before we proceed, there is one remaining class of constraints to address.
We encoded analyticity in the fixed-$t$ dispersion relation (6), which is $s\leftrightarrow u$ symmetric, but lacks $t\leftrightarrow u$ crossing symmetry constraints:
\be
\mathcal{\vec F}(s,t) = \vec{M}(s,t) - C_{tu}.\vec{M}(s,4-s-t)=0 \label{xsing0}
\ee
In order to extract a discrete number of constraints from the last equation, 
we plug   the dispersion relation \reef{fixedtoN} for $M^I$ in the crossing equation \reef{xsing0},  next expand into partial waves, 
and finally  take a number of derivatives $\partial_s^n \partial_t^m$  at $ s, t=4/3$. We are left with
\be
{\cal F}^{(I)}_{n,m} \equiv  \int_{4}^\infty dv \sum_{\ell, J}^\infty    F^{(IJ)}_{n,m;\ell}(v) \, n_\ell \, \text{Im} f^{(J)}_\ell(v) = 0
\label{crossing_constraints_on} 
\ee
with 
$ F^{(IJ)}_{n,m;\ell}(v) {=} \partial_t^n \partial_s^m   F^{(IJ)}_\ell(v, s,  t)\big |_{ s, t=\tfrac{4}{3}}$, $n_\ell{=}16\pi(2\ell{+}1)$. 
The exact form of  $F_\ell^{(IJ)}$ follows from projecting the Kernels in \reef{fixedtoN} into partial waves. On a first reading of this note, its exact details are not too important  to follow the logic flow.  We note that the    lowest non-trivial constraint is for $(n,m)=(1,3)$ derivatives. 
For  instance, for the $(I,J)=(3,1)$ channel we have
 $
F_{1,3;\ell}^{(31)}(v)= \frac{ (l+1) (x-1)^3 \left[\left([l+2] x^2-l\right) P_\ell(x)-2 x P_{\ell+1}(x)\right]}{   x^3 (x+1)^2}
$, with $x=\frac{4-3 v}{12-3 v}$, and 
 $P_\ell$ are Legendre polynomials. 
Similarly, higher order derivatives give rise  to   functions $F_{n,m;\ell}^{(IJ)}(v)$ consisting of linear combinations of Legendre polynomials times  rational functions of $v$ and $\ell$. The constraints in  \reef{crossing_constraints_on} are equivalent to the null constraints \cite{Caron-Huot:2020cmc,Tolley:2020gtv}.

\subsection{Rigorous dual bounds}

Our task  now is to find the extremal  values of $c_H$ under the constraints of unitarity, analyticity and crossing symmetry.
We will adapt to our needs the rigorous setup developed in~\cite{Guerrieri:2021tak}.~\footnote{The dual approach to the Bootstrap was first revisited in two dimensions~\cite{Cordova:2019lot}, and later generalized to scattering of several spiecies~\cite{Guerrieri:2020kcs}, and flux tubes~\cite{EliasMiro:2021nul}. In higher dimensions dual bounds were studied already long ago in~\cite{Lopez:1974cq,Lopez:1975wf,Lopez:1975ca,Bonnier:1975jz,Lopez:1976zs}.
A different  approach based on the Mandelstam representation was developed in~\cite{He:2021eqn}.}
With all the constraints laid down, an optimisation problem is best summarised by means of  a Lagrangian
\bea
&L_\pm(\{P\},\{D\})=
 \pm \,  c_H  
+{\color{gray} \underbrace{\black \sideset{}{_{(n,m)=(1,3)}^{(n_c,m_c)}}\sum
\nu_{n,m}^{(I)}{\cal F}_{n,m}^{(I)}  \hspace{-.2cm}\phantom{\Bigg|} }_{ s\leftrightarrow t \text{ crossing}}  } \label{lag} \\[-.1cm]
&\hspace{-.08cm}
+{\color{gray}\underbrace{\black \int_{4}^\infty  dv\,   \lambda^{(I)}_\ell(v) \, {\cal U_\ell}^{(I)}(v) \hspace{-.07cm}\phantom{\Bigg|}}_{\text{unitarity}}}
{+}{\color{gray}  \underbrace{\black\int_{4}^{\mu_c} dv\,   \sideset{}{_{j=0}^{J_c} }\sum \om^{(I)}_j(v) \, a_j^{(I)}(v) \hspace{-.2cm}\phantom{\Bigg|} }_{\text{analiticity + } s\leftrightarrow u\text{ crossing}} }  \nonumber
\eea
where repeated indices $I$ and $\ell$ are summed over,
and $a_j^{(I)}(v)$ is the $j$th partial wave projection of ${\cal A}^I(s,t)$.

 The first term in \reef{lag} is the objective to optimize.
The next terms  encode the crossing symmetry \reef{crossing_constraints_on},  unitarity \reef{xsing0},  and analiticity  \reef{fixedtoN} constraints  by means of the Lagrange multipliers \{$\nu_{n,m}^{(I)}$, $w_J^{(I)}$, $\lambda_\ell^{(I)}\geq 0$\}.  Collectively, all the Lagrange multipliers are denoted as \emph{dual variables} $\{D\}$.
 The primal variables $\{P\}$ are given by $\{\text{Re}f_\ell^{(I)}(s),\text{Im}f^{(I)}_\ell(s),c_H, c_\lambda\}$.

We keep a finite number   of crossing ${\cal F}_{m,n}$'s and spin projections $a_j(s)$'s  constraints --  the maximal  number is labeled by $(n_c,m_c)$ and $J_c$ respectively. Similarly we keep a maximal value $\mu_c$ in the evaluation of the projected dispersion relation ${\cal A}$. 
Even-though this is enough  for the derivation of a rigorous bound, we are   not leveraging the  full power of all the constraints that  we know of.
However  we will argue that the  observable that we are studying converges rapidly in $(n_c,m_c)$, $J_c$ and $\mu_c$.  Therefore, despite of truncating the number of constraints, our bounds will be close to optimality. 

Say we are interested in maximising $c_H$ in \reef{lag} (i.e. we take $+$ in the first term). The \emph{weak duality theorem}~\cite{sfi,*mathbook2} states that
\be
c_H\leq d_+(\{D\})  \equiv    \underset{\{P\}}{\text{max}}  \,  L_+(\{P\},\{D\})
\label{weakduality}
\ee
The maximization over the primal variables is straightforward since the Lagrangian~\eqref{lag} is a quadratic function. 
Below we summarize the features of the dual problem, further details can be found in appendix~\ref{dfder}.

 First, we discuss the maximization of $L$ w.r.t. $c_\lambda$ and $c_H$. Since they enter linearly into the Lagrangian as $\alpha c_i$, by taking derivatives we get that $\alpha$ must vanish, yielding the two normalization conditions
 \bea
 c_\lambda:  & \ \ 0= \int_4^{\mu_c} \frac{dv}{16 \pi} \, \vec{n}_1\cdot \vec N(v)  \label{clcon}  \\
 c_H: &  \ \  1= \int_4^{\mu_c} \frac{dv}{16 \pi}\left[ (v-\tfrac{4}{3}) \vec n_2  +(v-4) \vec n_3 \right]  \cdot \vec N(v)  \label{chcon}
 \eea
 where $\vec n_1=(n+2,2,0) $, $\vec n_2= (n-1,-1,0)$, $\vec n_3= (0,0,1/3)$ and $\vec N=(\om_0^1,\om_0^2,\om_1^3)$.
Maximizing the Lagrangian \reef{lag} with respect to $\{\text{Re}f_l^{(I)}, \text{Im}f_\ell^{(I)}\}$ and substituting the equations of motion leads to
 \be
 d_+(\nu,\lambda,\omega) =  \int_4^\infty \frac{dv}{\lambda_\ell\rho}   \left(\frac{n_\ell\mu_\ell}{2}+\lambda_\ell\right)^2+ \int_4^{\mu_c} \frac{dv }{4\lambda_j\rho} \,  \om_j^2  \label{dfunc}
 \ee
where we left implicit a sum on the channels $I$ and repeated indices are summed over according to \reef{lag}. 
\be
\mu_\ell^{(I)} =  \nu^{(K)}_{n,m} \cdot   F^{(KI)}_{n,m;\ell}(v)  - \text{p.v.} \int_4^{\mu_c} w_j^{(K)} k_{j,\ell}^{(KJ)}(s,v) 
\label{eq:muell}
\ee
where repeated  indices $\{K,n,m,j\}$  are summed over,   the $j$ and $(n,m)$ sums are cut according to \reef{lag},  `p.v' denotes the Cauchy principal value, and $\ell=0,1, \dots \infty$.  The kernel $k_{j,\ell}^{(KJ)}$ is given  in   appendix~\ref{disprel}.

For any value of the multipliers the inequality holds $ d_+(\nu,\lambda,\omega)\geq c_H$, and we obtain a bound on $c_H$. 
Thus, to obtain the best bound we should minimize $d_+$ over the Lagrange multipliers.  In practice it is hard to perform such minimization    analytically. 
Nevertheless, an efficient numerical algorithm to search for the minimal value of $d_+$ in the $\nu, \lambda, \omega$ space was developed in~\cite{Guerrieri:2021tak}. 
The generalisation to our problem is explained in  appendix~\ref{appni}.

\section{The space of \emph{O(4)} theories}

\label{island}

 \begin{figure}[t] \centering
 \includegraphics[scale=.45]{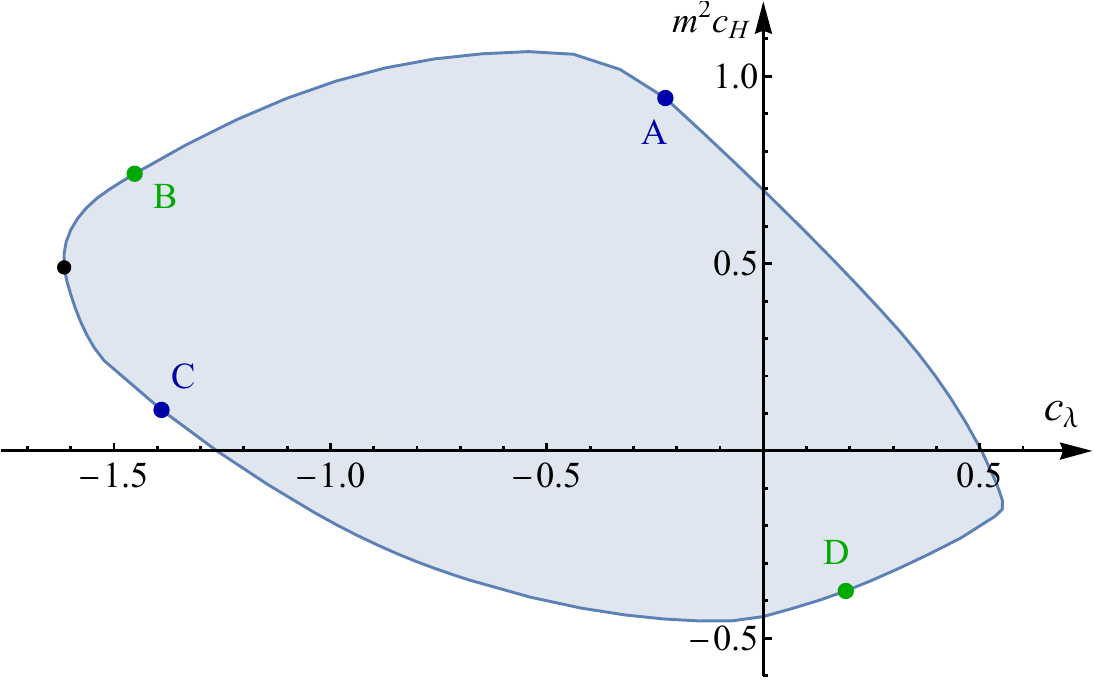} 
 \caption{All $O(4)$ theories must lie inside the  coloured region.   }
  \label{fig:dual_primal_oyster} \end{figure}

\subsection{The $O(4)$ non-perturbative island}
\label{islando4}

Our first goal is to determine universal bounds on $c_\lambda$ and $c_H$ defined in \eqref{amp}. 
By universal we mean that we are not making any assumption beyond the rigorous analyticity, crossing, and unitarity properties~\cite{Martin:1965jj}.
For concreteness we set $c_\lambda=R\cos\theta$, $c_H=R\sin\theta$, and for each fixed $\theta$ we bound the maximum value of $R$.
Our numerical results are shown in fig.~\ref{fig:dual_primal_oyster} -- see appendix~\ref{appni} for detailed explanations on the numerics. 
 Everything except  for the blue region is rigorously excluded:
all $O(4)$ theories must take values inside the blue `$O(4)$ island' in fig.~\ref{fig:dual_primal_oyster}.

The boundary of the island is determined by the extremal values that $c_\lambda$ and $c_H$ can take.
As we are not making any assumption, it is natural to expect that our bounds will be saturated by strongly coupled amplitudes all the way to the IR. A signature of strongly coupled IR dynamics is the presence of bound states or resonances. We experimentally observe the presence of scalar threshold bound states in the spin zero singlet and/or in the symmetric channel. Using this knowledge, we can define four distinct regions on the boundary of the island, whose properties are summarized in Table~\ref{tab}.

For instance, in the region $DA$ we have both the singlet and the symmetric threshold bound states. On the other hand, in region $BC$ there are none. However, even without threshold bound states there are other strong coupling phenomena happening. 
Between the point $B$ and the point with minimum value of $c_\lambda$ denoted by a black dot, although the value of $c_H$ is positive, we measure a negative spin one scattering length in the antisymmetric channel. This change of sign cannot be realised with a weakly coupled field theory description. On the other hand, between the black dot and point $C$, we find a spin one resonance at low energies.
In appendix~\ref{psapp} we have included a number of plots showing   the  phase-shifts of the amplitude around
the boundary of fig~\ref{fig:dual_primal_oyster}.

\begin{table}[t]
\centering
\begin{tabular}{ l   c   c  c  c  c } 
& \ A-B  \ &   \  B-C  \ &  \ C-D  \ &  \ D-A \   \\[0.4ex] 
 \hline 
 \hline
 singlet & $\checkmark$ & \ding{55}  & \ding{55} & $\checkmark$ \\
  \hline
 symmetric & \ding{55}  & \ding{55} & $\checkmark$ & $\checkmark$ \\
 \hline
\end{tabular}
\caption{Threshold singularities along the boundary.}
\label{tab}
\end{table}

\subsection{Perturbative boundary regions}
\label{pbr}

There are two linear combinations of dimension-eight operators of the $O(n)$ theory \reef{amp} that are positive~\cite{Adams:2006sv}: 
\bea
c_2&=\frac{1}{\pi}\int_4^\infty \frac{dv}{\bar v^3}\Im \vec{M}(v) \cdot \left( 0,\frac{1}{2},\frac{1}{2} \right) \geq 0, \label{c2sumrule}\\
2c_2+c_2^\prime&=\frac{1}{\pi}\int_4^\infty \frac{dv}{\bar v^3} \Im \vec{M}(v)\cdot \left( \frac{1}{n} , \frac{n{-}1}{n},0 \right) \geq 0.\label{c2primesumrule}
\eea
Both inequalities are saturated when the theory is free. Therefore, scanning the values of $c_\lambda$ and $c_H$ in the region where these two linear combinations are small, we single out  weakly coupled extremal amplitudes at the boundary of the allowed region. An example of this, is shown in fig.~\ref{bowls}.  The red line is analytically computed by performing a one-loop computation in $\lambda |\vec \phi|^4$, choosing the scheme in which $\lambda=c_\lambda$, and plotting the parametric curve  $\{c_H(c_\lambda),c_2(c_\lambda)\}$ -- in  appendix~\ref{lecs}  these functions are given.
The red line agrees with the boundary of the allowed region for  small $c_\lambda$. Interestingly, we do not have a perturbative description of the whole region around the origin. The boundary is expected to be  saturated by  amplitudes obtained from integrating out strongly coupled UV dynamics.~\footnote{ It is possible to ask   several other variations of the questions that we have  asked so far.   E.g. one could  min./max.  $c_H$  a a function of $\{\alpha,\beta\}$, with   $\alpha\equiv  \text{max}(c_2)$ and $\beta\equiv  \text{max} (2c_2+c_2^\prime)$. 
Small values of   $\{\alpha,\beta\}$ isolate  perturbative  amplitudes.
}

\subsection{EFT bounds}
\label{eftb}

A key property of EFTs is the  scale separation between the mass of the  scattered particle and 
the scale of `new physics' $\Lambda^2\gg m^2$. 
To incorporate the separation of scales non-perturbatively it is useful to introduce the concept of \emph{UV/IR domination} of the sum rules.
Consider the dispersive representation of $c_2$  split into  two pieces
\be
c_2^\text{IR} {=}  \int_4^{\Lambda^2}\hspace{-.2cm} \frac{dv}{\pi\bar v^3}\Im \vec{M}(v) \cdot\left( 0,\frac{1}{2},\frac{1}{2} \right) \, \text{,}\,\,\,\, c_2^\text{UV}{=} c_2-c_2^\text{IR}.
\ee
If $c_2^{IR}\gg c_2^{UV}$, then 
the sum rule is \emph{IR dominated}.
Conversely, if $c_2^{IR}\ll c_2^{UV}$ the dispersive integral receives the largest contribution from values at $s\gtrsim \Lambda^2$, in which case we say it is \emph{UV dominated}. In the case of $c_H$ this separation is less universal because of the explicit subtraction term. However, being IR or UV dominated is a physical property of the amplitude, not just of the sum rule. In the case of $c_H$ we will apply this definition to the dispersive part of the sum rule.

 \begin{figure}[t] \centering
 \includegraphics[scale=.55]{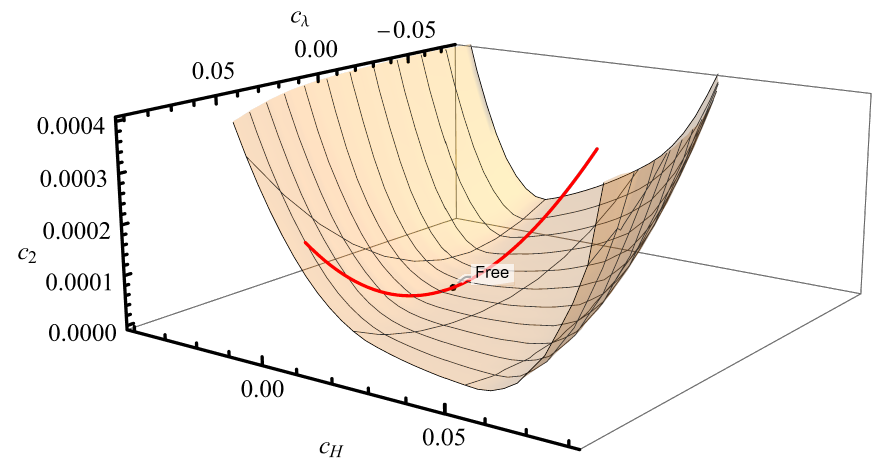} 
 \caption{A portion in the  boundary of  $O(4)$  theories that can be described by perturbative QFT.  }
  \label{bowls} 
\end{figure}

The bounds discussed in \ref{island} and \ref{islando4} are IR dominated with $\Lambda^2-4m^2=\mathcal{O}(1)m^2$ either because the amplitude is strongly coupled at values of $s \gtrsim m^2$ (fig.~\ref{fig:dual_primal_oyster})
or because the amplitude is weakly coupled at all energies with no significant resonance behaviour (fig.~\ref{bowls}).
In either case there is no effective separation of scales between $m^2$ and a putative `new physics' scale $\Lambda^2$.

Next, we are interested in the other limit, i.e. theories that  are  
 fully UV domi\-nated. 
 These are EFTs that are so weakly coupled in the IR $s\leq \Lambda^2$, 
 that the IR contribution to the dispersive integrals is negligible with respect to the UV contribution.~\footnote{Many recent interesting developments \cite{Arkani-Hamed:2020blm} exploiting positivity constraints apply in this regime.  }
 In this scenario, the sum rule \reef{chsr} gets replaced by 
 \be
\frac{\pi}{3} c_H s = \text{Re}f^{(3)}_1(s) - \int_{\Lambda^2}^\infty dv  \, k^{(3,J)}_{1,\ell}(s,v)  \text{Im}f^{(J)}_\ell(v) 
\label{chsrtwo}
\ee
where ignore powers of $m^2$ because we are considering $\Lambda^2 \gg m^2$.
 In~\cite{EliasMiro:2022xaa} it was shown that the bounds obtained in this regime are valid even in presence of a small physical IR imaginary part, which can be incorporated into a systematic error on the bound itself. The smaller the IR physics, the better is this approximation.

The UV contribution is not necessarily strongly coupled for UV domination to hold. For instance the whole amplitude may be well approximated at tree-level at all energies, but the exchange of a tree-level resonance localises with a delta function ($\text{Im}(s-\Lambda^2+i \eps)^{-1}\propto \delta(s-\Lambda)$) the UV integral at $s=\Lambda^2$.~\footnote{This is often the case for large N QCD-like theories~\cite{Albert:2022oes,Fernandez:2022kzi}.}

Next, we find the min/max values of $c_H$.
We do so by neglecting the imaginary part of the amplitude at values $s < \Lambda^2$  and by taking the massless limit $m{\to} 0$.
The procedure is a simple modification of what we described in section~\ref{space} and thus details are  relegated to  appendix~\ref{dfder}. 
There is  a single scale in the problem $\Lambda^2$,  and therefore bounds on $c_H$ are naturally  expressed 
by normalising with respect to $\Lambda^2$.
We will be  interested in  looking for a field theory interpretation of the bound, therefore we set $c_\lambda \ll 1$.
We call the theories showing UV dominated sum rules \emph{UV dominated EFTs}.

All in all, we find the result in \reef{bound2}.
This is a universal bound to all UV dominated theories: as long as  $m^2 \ll \Lambda^2$
and the dispersive part of $M(s|t,u)$ is negligible for $s< \Lambda^2$,  any such theory should satisfy the bound!

\begin{figure}[t] \centering
 \includegraphics[scale=.6]{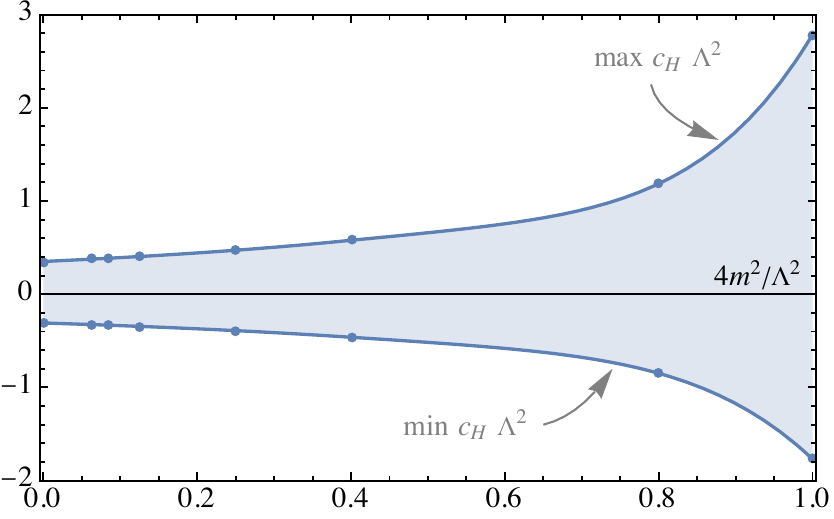} 
 \caption{Allowed value of $c_H\times \Lambda^2$ as a function of $4m^2/\Lambda^2$.   }
  \label{fig:interpolation} \end{figure}

For completeness it is interesting to interpolate between the  UV and IR domination regimes. 
We do so by min/max $c_H$ defined through   
$
\frac{\pi}{3} c_H (s{-}4m^2) 
= \text{Re}f^{(3)}_1(s) - \int_{\Lambda^2}^\infty dv  \, k^{(3,J)}_{1,\ell}(s,v)  \text{Im}f^{(J)}_\ell(v) 
$
neglecting the imaginary part of the amplitude in $[4m^2, \Lambda^2]$ and varying $\Lambda$ within $[4m^2, \infty)$.
The result of this exercise is given in fig.~\ref{fig:interpolation}.
The rightmost points correspond to the min/max values of $c_H$ in the $\frac{4m^2}{\Lambda^2} \rightarrow 1$ limit. 
Those points agree with fig.~\ref{fig:dual_primal_oyster}, at $c_\lambda=0$, after accounting for the  $\frac{4m^2}{\Lambda^2}$ normalisation factor, $4\times[-0.44,0.70]=[-1.76,2.8]$. 
The leftmost points are instead the UV domination limit $\frac{m^2}{\Lambda^2} \rightarrow 0$, in agreement with \reef{bound2}.
Note that for $\frac{m}{\Lambda} \lesssim \frac{1}{8}$ the bound  is close to the asymptotic bound   $\frac{m^2}{\Lambda^2} \rightarrow 0$ 
and shows variation only below the percent level.

One might ponder how the just-derived boundary will be influenced by the introduction of a small non-vanishing discontinuity in the IR. This matter is discussed in the next section

\subsection{Rigorous bounds assuming an IR model}
\label{IRmodel}

In this section, we introduce a small IR imaginary part and study its effect on our dual bounds.
We address this question using the following model. We imagine that somebody gives us a functional fit of the imaginary part of the amplitude for all spins and irreps below a certain energy scale $\Lambda$
\be
\text{Im} f_\ell^{(J)}(s)\equiv g_\ell^{(J)}(s),\quad s<\Lambda^2.
\label{eq:low_energy_model}
\ee
We can now extremize the values of $c_H$ combining the set of constraints in~\eqref{lag} with the new conditions~\eqref{eq:low_energy_model}.
The generalization of  the dual problem to include this additional constraint is straightforward and is discussed in appendix~\ref{appdir}.
The final dual bounds take the form
\be
D_-^{UV}+D_-^{IR}[g^{(I)}_\ell]\leq c_H\leq D_+^{UV}+D_+^{IR}[g^{(I)}_\ell]
\ee
where $D^{UV}$ is the functional used to obtain the EFT bounds shown in fig.~\eqref{fig:interpolation}, and $D^{IR}[g^{(I)}_\ell]$ is the IR contribution which vanishes if $g^{(I)}_\ell=0$. For concreteness, if we take $\Lambda=8m$, and choose 
\bea
g^{(I)}_0= \frac{\lambda^2/2}{(16\pi)^2} \sqrt{\frac{s-4}{s}} \, \big( (n+2)^2, 4, 0 \big) , \quad g^{(I)}_{\ell>0}=0
\eea
with $\lambda=0.1$, we obtain 
\be
-0.33 < c_H\times 8^2 <0.38.
\ee
The difference between this bound and the one obtained by neglecting the imaginary part in fig.~\eqref{fig:interpolation} is of order $10^{-5}$ for this value of the quartic coupling. 

In this  analysis we have not included nonlinear unitarity in the IR. If included, it would be interesting to   compare with   the bounds in \cite{EliasMiro:2022xaa}, which were obtained by solving the primal problem.
We leave this exploration to a future work.

%
%

\section{Dimension-six operators}
\label{dim6}

The bounds on $c_H$ that we have presented thus far are sharp and rigorous. 
Moving forward, next we aim to interpret them through an effective field theory Lagrangian.
While this will necessitate making additional assumptions, it will also enable us to make further predictions.
Once we establish a match between the amplitude's coefficient, denoted as $c_H$, and the effective operator $|\partial H|^2 |H|^2$, it opens up new opportunities to test the constraints on $c_H$. 
Besides altering high energy $2\rightarrow 2$ scattering of  the Higgs particles  or longitudinal Electroweak gauge bosons, 
this operator  universally  modifies of all Higgs couplings. 
Indeed, after accounting for the Higgs vacuum expectation value it leads to a Higgs wave function re-normalization. 
Thus, the interpretation of our bound on $c_H$ in terms of  the  field theory operator allows us to determine the maximal deviation on Higgs couplings due to the $O_H$ operator. 
For instance $\Gamma(h\rightarrow VV)_\text{SILH}/\Gamma(h\rightarrow VV)_\text{SM}=1-(v/f)^2 c_H+ \cdots$, where $\cdots$ denote  other  Wilson coefficients~\cite{Giudice:2007fh}.
Thus, even though the  dimension-six operator is not a clean observable,  establishing a connection with the bound on $c_H$ is a worthwhile exercise due to the physics motivations just explained.

\subsection{Maximally UV dominated EFTs}
\label{bdso}

Consider the field theory given by 
\be
{\cal L}_H= {\cal L}_\text{Free}- \epsilon |H|^4 - \frac{g_H}{f^2} |\partial H|^2 |H|^2 +{\cal O}(f^{-4}) \label{inteft}
\ee
with $\eps=m^2/f^2 \ll 1$. 
A simple calculation  of the amplitude \reef{amp} reveals $c_\lambda=O(\eps)$ and  $c_H \times \Lambda^2 =\frac{g_H}{(4\pi)^2}\frac{\Lambda^2}{f^2}{+}{\cal O}(\eps)$. Higher order coefficients do receive  possibly large corrections from $g_H$, e.g. $c_2 \times \Lambda^4 =g_H^2 \frac{\Lambda^4}{f^4}+ \dots  $. 
Note also that $\text{Im} M={\cal O}(\eps^2,\epsilon g_H)+g_H^2\times {\cal O}( s^2/f^4)$, and therefore 
in the perturbative  computation of the two-to-two scattering amplitude
it is justified to neglect the imaginary part of the amplitude provided that  $s$ is well below the value of  $f^2/g_H$.

In the extreme UV domination limit, and when the 
 gap is large $\eps^{-1} \gg 1$, equation   \reef{bound2} implies the  bound on the dim.-six operator coeffi\-cient 
\be
-0.31  < \frac{g_H}{(4\pi)^2}  \times  \frac{\Lambda^2}{f^2}  < 0.35     \, . \label{Qdim}
\ee 
\emph{What is  the appropriate value of $\Lambda^2/f^2$?}

\noindent  The scale $\Lambda^2$ was introduced to  ensure that  the IR contribution to the $c_H$ sum rule \reef{chsr} is  small with respect to the UV contribution.
 Thus the derivation of   \reef{Qdim}   is valid as long as     
\be
\hspace{-.08cm}\int_{0}^{\Lambda^2} \hspace{-.2cm} dv  \,   \kappa^{(J)}_\ell(v)  \,  \text{Im}f^{(J)}_\ell(v)   \ll  \int_{\Lambda^2}^\infty \hspace{-.2cm} dv  \,   \kappa^{(J)}_\ell(v)  \,  \text{Im}f^{(J)}_\ell(v)  \label{tuning}
\ee
which  follows from  \reef{chsr} in the limit $\eps \rightarrow 0$, and we have defined $\kappa^{(J)}_\ell(v) \equiv k^{(3,J)}_{1,\ell}(0,v) $.
Instead the scale $f^2$ does not have an intrinsic definition within the EFT.~\footnote{
Given the value of $g_H$,  $f^2$ is often associated to the lowest energy scale at which  the perturbative calculation of $2\rightarrow 2$ breaks down.
However this definition would be somewhat circular and not useful to us. 
The other standard  choice is to associate  $f^2$  to the scale of new physics, where the EFT breaks down. 
This definition is not useful for our purposes either.}
Thus we   define $f^2$ as  the scale at which \reef{tuning} is satisfied for the largest~\footnote{Equation \reef{tuning} is trivially satisfied if $\Lambda^2$ is taken arbitrarily small.} possible $\Lambda^2$. 
That is we will take $f^2 = \Lambda^2$.

We remark that we are not claiming a regime such that the imaginary part of the amplitude is  necessarily negligible at energies $s\lesssim f^2$. We are   instead arguing for the existence of a regime such that the IR contribution to the sum rule is subdominant with respect to the UV contribution \reef{tuning}.
After identifying $f^2=\Lambda^2$, the  remaining question is
for what type of theories the condition being assumed  \reef{tuning} 
is less constraining than the actual result \reef{Qdim}. 
While we do not know the answer to this question in its most general terms, 
next we will  provide two sources of intuition. 

The first one comes from   simple perturbative models. 
As argued above, the effect of  exchanging heavy weakly coupled resonances on the dispersion relation is to localise the dispersive integrals at the heavy particle threshold. 
Thus, if the IR couplings are parametrically smaller than the UV couplings to heavy states, then UV domination \reef{tuning} follows.
 As the UV couplings becomes stronger  the condition \reef{tuning}
   still  holds if the IR couplings are hold  weaker.
   A simple perturbative  example full filling  this behaviour of UV/IR domination  is worked out on appendix~\reef{apptun}.
   
    \begin{figure}[h] \centering
 \includegraphics[scale=.75]{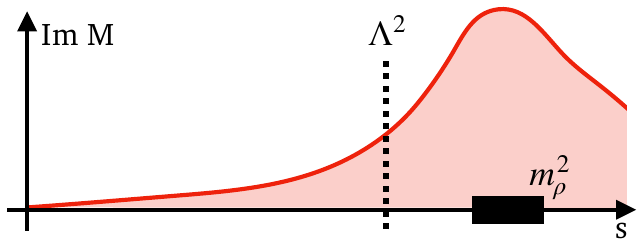} 
 \caption{Representation of strongly coupled and UV dominated  amplitudes.
 The amplitude is strongly coupled at energies $s\gtrsim \Lambda^2$, nevetheless \reef{tuning} is satisfied, i.e. the $c_H$ sum rule is UV dominated.   }
  \label{uvir} \end{figure}

  Even if the scattering amplitude can not be computed in perturbation theories for energies $s\leq\Lambda^2$, 
  the condition \reef{tuning} may still be satisfied. For an intuitive picture see figure~\ref{uvir}.
In our previous work~\cite{EliasMiro:2022xaa} we  constructed scattering  amplitudes meeting this trait,
as well as   amplitudes that  smoothly interpolate between the UV and IR domination regimes -- providing our second source of intuition.  
These theories show broad resonance behaviour for  $s >\Lambda^2$, with large values of the imaginary part for $s>f^2$.  For energies below the resonances, the amplitude decays in powers of energy over the scale of the new resonances.

Accidentally  equation \reef{Qdim} with $\Lambda^2=f^2$  agrees with the rough   `loop-democracy'  estimate -- often called Naive Dimensional Analysis (NDA) ~\cite{Manohar:1983md,*Georgi:1986kr,*Luty:1997fk}. 
Indeed, using \reef{inteft} to compare   tree v.s. one-loop corrections to the four-point function gives $g_H \gtrsim g_H^2/(4\pi)^2 s/f^2 $. 
 We are conservatively arguing for  $s\sim \Lambda^2 =f^2$, and are lead to  the bound $|g_H|\leq {\cal O}(1) \times (4\pi)^2$ in \reef{Qdim}. 
What we have achieved here is to turn the NDA  estimate 
into a precise `theorem'  by   determining the
order one factors.
 The bound we have uncovered shows the symmetry $|g_H^\text{max}|\approx  |g_H^\text{min}|$, a  trait  that was hardly predictable prior to the calculations presented here.~\footnote{Indeed such symmetry is absent for the closely related bound \reef{bound1}.}

 Various composite Higgs models, where the light Higgs is a pseudo-goldstone boson,  have been shown to fall in within the SILH  power-counting~\cite{Giudice:2007fh}. For instance  the holographic Minimal Composite Higgs Model~\cite{Agashe:2004rs} gives $g_H=1$.
  These type of models are well within \reef{Qdim}.
 Our bound could be  made more stringent with further understanding or assumptions about  the extend to which UV domination \reef{tuning} holds for these particular class of models.  For instance one could argue to improve the bound by  pushing $\Lambda^2$ to larger values by  setting $\Lambda^2  \equiv  f^2 g_\rho^2 \equiv  m_\rho^2$. 
 A more interesting  possibility is to improve our bound by   further  modelling of the IR, in the spirit of section~\ref{IRmodel}.  Besides of requiring a separation of scales with  weak coupling in the IR \reef{tuning}, models where the Higgs arises as a pseudo-goldstone boson can be further  characterised by imposing chiral zeros on the scattering amplitude, very much like in pion physics~\cite{Guerrieri:2018uew}.  We leave this intriguing possibility to improve the bound for these class of theories to future investigations.

Defining $f^2$ in terms of $\Lambda^2$ appears to us as the only logical possibility  for establishing rigorous bounds on dim.-six operators. 
Given that the SM is very weakly coupled at TeV energies, it is  reasonable   to associate   $\Lambda$ ($\equiv f$)
with the largest energy scale for which  new physics, or new resonances, have been excluded. 
Namely, to the extend that no new physics contributes to the left hand side (l.h.s.) of  \reef{tuning} up to an energy scale $\Lambda$, 
it is safe to neglect it~\footnote{ For UV completions above $\Lambda$ that are parametrically weaker than IR physics, the l.h.s. is not negligible. However bounds  for these   theories are of little use  and  we shall not consider them furhter.}. 
Therefore, if no new physics appears up to the scale $f^2\equiv \Lambda^2 $, then \reef{Qdim} is the maximal value of $g_H$ that one can hope of measuring. 
The current fit to the LHC data reveals     $|g_H| \leq 1\times \Lambda^2/(\text{1~TeV})^2$~\cite{Ethier:2021bye}.
Our bound is universal in units of the scale  $\Lambda^2$ (recall that as long as $\Lambda\gtrsim 8 m$, we are on the asymptotic left region of fig.~\reef{fig:dual_primal_oyster}).
Thus, if we assume no new physics enters on the r.h.s. of  \reef{tuning} up to an energy scale  $5~\text{TeV}$, then our bound reads
$
-0.31 \times (4\pi)^2/ 5^2
  < g_H   < 0.35    \times (4\pi)^2/ 5^2
$, i.e. 
$
-2.0
  < g_H   < 2.1 $ which is comparable to the current experimental bound. 
  Our construction can thus be used to figure out the precision needed on Higgs coupling  measurements  given an exclusion bound on the energy scale of new physics.


\section{Future directions}
 \label{so}
 
This program is in its early stages. 
We presented the answer to a very specific problem, what is the maximal/minimal value of $c_H$  and where do SM-like EFTs fit within this bound. 
We are lead  to many more questions that would be interesting to investigate, next we present few of them:
 \begin{itemize} 
   \item  An interesting aspect of the starting point we took is that it can be  extended to other theories and make it more realistic. 
   Further modelling of the IR amplitudes will allow for more refined bounds. This modelling is biased on the particular IR physics one is interested in probing, which is why in this work we have restricted ourselves to  fairly simple choices. 
   \item  It would be interesting to constrain the dimension-six operators  involving two $H$'s and two electroweak gauge bosons.~\footnote{
The  are five    operators $D_\mu W^{a\, \mu\nu}H^\dagger \sigma^a D_\nu  H$,  $D_\mu B^{\mu\nu}H^\dagger D_\nu H$, 
$H^\dagger \sigma^i H W^{i\mu\nu}B_{\mu\nu}$,  $ |H|^2 W_{\mu\nu}W^{\mu\nu}$, $ |H|^2 B_{\mu\nu}B^{\mu\nu}$.}
These operators, together with $c_H$, control  the deviation of $h\rightarrow W^+W^-/ZZ$ within the SILH framework. 
 \item  Our bound \reef{Qdim} is fairly symmetric, $|g_H^\text{max}|\approx  |g_H^\text{min}|$. It would be phenomenologically interesting to know whether other dimensions-six operators   enjoy instead very asymmetric bounds. 
  \item Related to the previous point, we remark  that  our bound is not optimal. 
 It is nevertheless sufficiently stringent to provide a rigorous constraint within the relevant experimental ballpark. 
The primal $S$-matrix Bootstrap~\cite{Paulos:2017fhb} is well suited to study in detail the physical properties of the UV completed amplitudes saturating the bounds studied. 
It would be interesting to close the gap between the UV dominated primal amplitudes constructed in \cite{Chen:2022nym,EliasMiro:2022xaa} and the dual bound approach here presented in~\ref{IRmodel} by including also nonlinear unitarity in the IR. 

 \item  As a final remark, there are many generalizations of our approach that could be done. 
From the dual perspective, it might be interesting to extend the dual problem including crossing symmetric dispersion relations, recently reviewed in~\cite{Sinha:2020win}, see also~\cite{Zahed:2021fkp,Li:2023qzs}, and check whether this helps reconstructing the phase shifts in a larger domain.
Another direction is to study the scattering of massless particles, assuming an IR EFT input, and compare, for instance, with the results obtained in~\cite{Acanfora:2023axz}. Finally, it would be also interesting to use the recursive approach of~\cite{Tourkine:2023xtu}, and reconstruct the full amplitude starting from an IR EFT input.

 \end{itemize}

 \section*{Acknowledgements}
 We thank Brando Bellazzini, Jo\~ao Penedones, Ric\-cardo Rattazzi, Marc Riembau, Francesco Riva, Rogerio Rosenfeld and Slava Rychkov for useful discussions. 
 JEM  is supported by the European Research Council, grant agreement n. 101039756.
MAG thanks PI for partial support during the completion of this project. 
 Research at the Perimeter Institute is supported in part by the Government
of Canada through NSERC and by the Province of Ontario through MRI. ALG is supported by the
European Union - NextGenerationEU, under the programme Seal of Excellence@UNIPD, project
acronym CluEs.



\small

\bibliographystyle{utphys}

\bibliography{biblio}



\clearpage
\appendix
\onecolumngrid

\section{$O(n)$ amplitudes and dispersion relations}
\label{disprel}

The two-to-two scattering amplitude of  scalars transforming as vectors under an $O(n)$ internal symmetry, $v_a+v_b\rightarrow v_c+v_d$,
 can proceed through three different channels, as shown in   \reef{anp}.
In this theory crossing symmetry amounts to the fact that the function  $M$ is  symmetric under the exchange of its last two variables $M   (\bar s | \bar t, \bar u)=M   (\bar s | \bar u, \bar t)$. 

Unitary acts diagonally when expressed in terms of the three irreducible representations of $v_a\otimes v_b$,  
respectively the singlet, symmetric, and antisymmetric irreps,
\bea
M^{(sing)} &= n \, M(\bar s | \bar t , \bar u) + M(\bar t | \bar u , \bar s) + M(\bar u | \bar s , \bar t)  \, , \\[.2cm] 
M^{(sym)}  &= M(\bar t | \bar u , \bar s) + M(\bar u | \bar s , \bar t) \, ,   \\[.2cm]
M^{(anti)} &= M(\bar t | \bar u , \bar s) - M(\bar u | \bar s , \bar t) \, . 
\eea
We will often group the irreps into the vector $ \overrightarrow{ M}\equiv (M^{(sing)} , M^{(sym)} , M^{(anti)} )$.
Then, the action of crossing symmetry on the irreps follows from their definition and 
the symmetry properties of $M$. 
It turns out that
\be
 \overrightarrow M(\bar s ,\bar t, \bar u)=C_{st}. \overrightarrow M(\bar t ,\bar s, \bar u)  \ , \quad
 \overrightarrow M(\bar s ,\bar t, \bar u)=C_{su}. \overrightarrow M(\bar u ,\bar t, \bar s) \ , \quad
 \overrightarrow M(\bar s ,\bar t, \bar u)=C_{tu}. \overrightarrow M(\bar s ,\bar u, \bar t) \, ,
\ee 
where the crossing matrices satisfy 
  $C_{st}.C_{su} = C_{tu}.C_{st} = C_{su}.C_{tu}$,
  $C_{tu}=\text{diag}(1,1,-1)$,  
  \be
\renewcommand*{\arraystretch}{1.5}
C_{st} =
 \left(
\begin{array}{ccc}
\frac{1}{n} & \frac{1}{2}{-}\frac{1}{n}{+}\frac{n}{2} & \frac{n}{2}{-}\frac{1}{2} \\ \frac{1}{n} & \frac{1}{2}-\frac{1}{n} & -\frac{1}{2} \\ \frac{1}{n} & -\frac{1}{2}{-}\frac{1}{n} & \frac{1}{2} \\\end{array}
\right)    \, ,\quad
C_{su} = \left(\begin{array}{ccc}
\frac{1}{n} & \frac{1}{2}{-}\frac{1}{n}{+}\frac{n}{2} & \frac{1}{2}{-}\frac{n}{2} \\ \frac{1}{n} & \frac{1}{2}-\frac{1}{n} & \frac{1}{2} \\ -\frac{1}{n} & \frac{1}{2}{+}\frac{1}{n} & \frac{1}{2} \\\end{array}
\right)    \, ,
\ee
and $C_{st}^2=C_{su}^2=C_{tu}^2=\mathds{1}$.

The partial wave decomposition can be carried independently on each of the three different channels 
\be
M^{(I)}(s,t) = \sum_{\ell, I}^\infty n_\ell \, f^{(I)}_\ell(s) P_\ell (1{+}\tfrac{2t}{s-4}) \, , \label{pwaveflavor}
\ee
where $I=sing,\, sym\, , anti$, or equivalently $I=1,2,3$.
As usual 
\be
f_\ell^{(I)}=\frac{1}{32\pi} \int_{-1}^{1} dz P_\ell(z) M^{(I)}(s,t(z)),
\ee 
with $t(z)=-1/2(s-4)(1-z)$ and $n_\ell \equiv 16 \pi (2\ell+1)$.
Singlet and symmetric channels are symmetric under the exchange of $t \leftrightarrow u$, therefore their odd spin partial wave projections vanish.
Similarly, even spin partial wave projection of the antisymmetric channel amplitude are zero. 
The sum over the indices $\{\ell, I\}$ in \reef{pwaveflavor} means $\{\ell \text{ even}, I=(sing,sym)\}$ and $\{\ell \text{ odd}, I=(anti)\}$.

\subsection{Deriving the $O(n)$ Dispersion Relation}
\label{on_dispersion}


Let us outline how to derive a double subtracted dispersion relation for the $O(n)$ amplitude inspired by Ref.~\cite{Roy:1971tc}, whose subtraction constants we choose to be $c_\lambda$ and $c_H$. We start by writing a contour integral counterclockwise around an analytic point $z=s$ for fixed-$t$
\be
\overrightarrow{M} (s,t) = \frac{1}{2\pi i}\oint dz \, \frac{1}{z^2} \frac{s^2}{z-s} \overrightarrow{M} (z,t) \, .
\ee
Then we blow up the contour to infinity, and by using $\overrightarrow{M}(s,t)=\frac{1}{2} \overrightarrow{M}(s,t) + \frac{1}{2} C_{su}.\overrightarrow{M}(u,t)$, we get
\bea
\overrightarrow{M}(s,t) &= \vec{f}(t;s,u) + \frac{1}{\pi} \int_4^\infty dz \, K(z;s,u). \overrightarrow{M}_z(z,t)
\eea
where the kernel $K$ is a matrix given by
\be
K(z;s,u)=\frac{1}{z^2} \left(  \frac{s^2}{z-s}  \mathds{1} +  \frac{u^2}{z-u} C_{su}\right)
\ee
satisfying $K(z;s,u) = C_{su}.K(z;u,s)$, and $\vec{M}_z \equiv \text{Disc}_z \vec{M}(z,t)/2i$, and $\vec{f}$ is a function that contains residues at $z=0$ and integrals along the left-hand cuts. Notice that the $s \leftrightarrow u$ symmetry of the amplitude $M$ and the kernel $K$ forces $\vec{f}$ to satisfy $\vec{f}(z;s,u) = C_{su}.\vec{f}(z;u,s)$. Below we replaced $\vec{f}$ with a simple ansatz with the desired $s \leftrightarrow u$ symmetry property, and obtained the following equations:
\bea
\overrightarrow{M}(s,t) &= C_{st}. \left[  \vec{c}(t) + (s-u)  \vec d(t) \right] + \frac{1}{\pi} \int_4^\infty dz \, K(z;s,u). \overrightarrow{M}_z(z,t), \label{dispN} \\
\partial_s \overrightarrow{M}(s,t) &= C_{st}. \left[ 2 \vec d(t) \right] + \frac{1}{\pi} \int_4^\infty dz \,\partial_s \left[ K(z;s,u)\right] . \overrightarrow{M}_z(z,t), \label{dispNder}
\eea
where $\vec{c}(t)=(c^\text{sing}(t),c^\text{sym}(t),0)^T$ and $\vec{d}(t)=(0,0,d^\text{anti}(t))^T$.
Evaluating \reef{dispN} and \reef{dispNder} at the crossing symmetric point, we can derive the following:
\bea
\vec{c}(\tfrac{4}{3}) &= (n+2,2,0)^T c_\lambda - \frac{1}{\pi} \int_4^\infty C_{st} \, . \, K(\tfrac{4}{3},\tfrac{4}{3}) \, . \, \vec{M}_z(z,\tfrac{4}{3})\\
2 \vec{d}(\tfrac{4}{3}) &= (0,0,2)^T c_H - \frac{1}{\pi} \int_4^\infty C_{st} \, . \, \partial_s K(s,\tfrac{8}{3}-s)|_{s=4/3} \, . \, \vec{M}_z(z,\tfrac{4}{3})
\eea
Then, using the equation $\vec{M}(4/3,t)=C_{st}.\vec{M}(t,4/3)$, we can reexpress $t$-dependent subtraction constants with the ones above
\be
C_{st}.\left[ \vec{c}(t) + (s-u) \vec{d}(t) \right] = \vec{c}(\tfrac{4}{3}) + 2(t-\tfrac{4}{3})\,\vec{d}(\tfrac{4}{3}) + 2(s-\tfrac{4}{3})\,C_{st}.\vec{d}(t) + \int absorptive,
\ee
as well as
\bea
2\vec{d}(t) = 2\vec{d}(\tfrac{4}{3}) + \frac{(\mathds{1}-C_{tu})}{(t-4/3)} \frac{1}{\pi} \int^\infty_4 dz \, \left[ K(t,\tfrac{8}{3}-t)-K(\tfrac{4}{3},\tfrac{4}{3}) \right].\vec{M}_z(z,\tfrac{4}{3}) - C_{st}.K(\tfrac{4}{3},\tfrac{8}{3}-t).\vec{M}_z(z,t). \nonumber
\eea
The integrand in the second term goes like $O((t-4/3)^2)$ so it is regular when $t \to 4/3$.
Finally, plugging everything back into \reef{dispN} yields
\be
M^{(I)}(s,t)=\mathbb{C}^{(I)}(s,t) + \int_{4}^\infty dz \, \left[ \, \mathbb{K}^{(IJ)}(z; s,t) \text{Im}\, M^{(J)}(z,t) +  \mathbb{L}^{(IJ)}(z; s,t) \text{Im}\, M^{(J)}(z,4/3) \, \right], \label{AfixedtoN}
\ee
where
\be
\vec{\mathbb{C}}(s,t)= c_\lambda \begin{pmatrix} n+2 \\ 2\\ 0 \end{pmatrix} + c_H \begin{pmatrix} n-1 \\ -1 \\ 1 \end{pmatrix} (s-4/3) + c_H \begin{pmatrix} 0 \\ 0 \\ 2 \end{pmatrix} (t-4/3).
\ee
Notice that the absorptive pieces vanish at $(s,t)=(4/3,4/3)$.
The explicit form of the kernels is given by:
\bea
&\mathbb{K}(z; s,t) = \frac{(3 s-4) (3 s+3 t-8)}{\pi  n (3 z-4) (3 t+3 z-8) (s+t+z-4)} \, \times \begin{bmatrix}
 1-\frac{n (s+t+z-4)}{s-z} & \frac{1}{2} \left(n^2+n-2\right) & -\frac{1}{2} (n-1) n \\
 1 & -\frac{n (s+2 t+3 z-8)}{2 (s-z)}-1 & \frac{n}{2} \\
 -1 & \frac{n+2}{2} & -\frac{n (s+2 t+3 z-8)}{2 (s-z)} \\
\end{bmatrix} \label{kernelK}  \\
\phantom{}\nonumber\\
&\mathbb{L}(z; s,t) = \frac{3 (n-1) (3 t-4)}{\pi  n (3 z-4) (3 t+3 z-8)} \, \times \label{kernelL} \\
&\begin{bmatrix}
 \frac{2 (3 t-4)}{3 (n-1) (z-t)}+\frac{4-3 s}{3 z-4} & \frac{3 n s-4 n+6 s-8}{2 (3 z-4)}+\frac{(n+2) (3 t-4)}{3 (z-t)} & \frac{n (2 s+t-4)}{z-t}-\frac{n (3 s-4)}{2 (3 z-4)} \\
 \frac{3 s-4}{(n-1) (3 z-4)}+\frac{2 (3 t-4)}{3 (n-1) (z-t)} & \frac{(n-2) (3 t-4)}{3 (n-1) (z-t)}-\frac{(n+2) (3 s-4)}{2 (n-1) (3 z-4)} & \frac{-2 n s-n t+4 n}{(n-1) (z-t)}+\frac{3 n s-4 n}{2 (n-1) (3 z-4)} \\
 \frac{2 (3 t-4)}{3 (n-1) (z-t)}-\frac{3 (s+2 t-4)}{(n-1) (3 z-4)} & \frac{3 (n s+2 n t-4 n+2 s+4 t-8)}{2 (n-1) (3 z-4)}-\frac{(n+2) (3 t-4)}{3 (n-1) (z-t)} & \frac{2 n s+n t-4 n}{(n-1) (z-t)}-\frac{3 (n s+2 n t-4 n)}{2 (n-1) (3 z-4)} \\
\end{bmatrix}. \nonumber
\eea

The $j$th partial wave projection of \reef{AfixedtoN} gives rise to the Roy equations~\cite{Roy:1971tc}
\be
a^{(I)}_j(s) = \text{Re} f^{(I)}_j(s) - \frac{\mathcal{C}^{(I)}_0}{16\pi}\delta_{0,j} - \frac{\mathcal{C}^{(I)}_1}{16\pi}\delta_{1,j} - \text{P.V.} \int^\infty_4 dv \sum^{\infty}_{\ell, J} k^{(IJ)}_{j,\ell}(s,v) \, n_\ell \, \text{Im}f^{(J)}_\ell(v) = 0 \, . \label{const_analyticity_on}
\ee
where the subtraction constants are given by
\be
\vec{\mathcal{C}}_0 = \begin{pmatrix} c_\lambda (n+2) + c_H (n-1)(s-4/3) \\ 2c_\lambda - c_H (s-4/3) \\ 0 \end{pmatrix} 
\quad , \quad
\vec{\mathcal{C}}_1 = \begin{pmatrix} 0 \\ 0 \\ (c_H/3)(s-4), \end{pmatrix} 
\ee
and the kernels by
\be
k^{(IJ)}_{j,\ell}(s,v) = \frac{2}{32\pi} \int_0^1 dz P_j(z) \left[ \mathbb{K}^{(IJ)}(v;s,t) P_\ell (1{+}\tfrac{2t(z)}{v-4}) + \mathbb{L}^{(IJ)}(v;s,t) P_\ell ( 1{+}\tfrac{8/3}{v-4} ) \right].
\label{eq:kernelsOn}
\ee
Attached to the arXiv submission the file \texttt{disp\_and\_plots.nb} provides the implementation of 
\reef{AfixedtoN}, \reef{eq:kernelsOn}, and few simple numerical cross-checks. 

\section{Dual functional}
\label{dfder}

Let us find $d_\pm$ defined in eq.~\eqref{weakduality}, corresponding to the problem of maximizing or minimizing $c_H$ using only rigorous analyticity, crossing, and unitarity. 
First maximizing $L_\pm$ w.r.t $c_\lambda$ and $c_H$, yields the following normalisation constraints
\be
\begin{aligned}
\label{clchcon}
\frac{\partial L_\pm}{\partial c_\lambda}=0:& & 0 &= \int_{\Lambda^2}^{\Lambda^2_c} \frac{dv}{16 \pi} \, (n+2,2,0) \cdot \vec{\omega}(v) \\
\frac{\partial L_\pm}{\partial c_H}=0:&   & \pm 1 &= \int_{\Lambda^2}^{\Lambda^2_c} \frac{dv}{16 \pi}\left[ (v-\tfrac{4}{3}) (n-1,-1,0) + (v-4) (0,0,\tfrac{1}{3}) \right]  \cdot  \vec{\omega}(v)
\end{aligned}
\ee
where $\vec{\omega} \equiv (w^{(1)}_0,w^{(2)}_0,w^{(3)}_1)$.
Next comes the maximization with respect to $\{\text{Re}f_l^{(I)}, \text{Im}f_\ell^{(I)}\}$. 
On their equations of motion
\begin{eqnarray}
\rho \cdot \text{Re} f_{(I\ell)} &=& \frac{w_{(I\ell)}} {2 \lambda_{(I\ell)}}, \\
\rho \cdot \text{Im} f_{(I\ell)} &=& 1 + n_\ell \, \frac{\mu_{(I\ell)}} { 2 \lambda_{(I\ell)}},
\label{dualpwaves}
\end{eqnarray}
where $\rho=\sqrt{(s-4)/s}$. After plugging in the solution, we get equation~\reef{dfunc} in full glory 
\be
d_\pm(\mathcal{D}) =
\int^\infty_{\Lambda^2} \frac{dv}{\rho} \cdot 
\frac{1}{4\lambda_{(I\ell)}} \left( 2\lambda_{(I\ell)} + n_\ell \, \mu_{(I\ell)} \right)^2
+
\int^{\Lambda^2_c}_{\Lambda^2} \frac{dv}{\rho} \cdot 
\frac{w_{(Kj)}^2}{4\lambda_{(Kj)}}
\label{dualwithlambda}
\ee
where we have grouped isospin and spin indices together in a common paranthesis for convenience, and we have defined an auxiliary function
\be
\mu_\ell^{(I)}(v) \equiv \nu^{(K)}_{n,m} \cdot F^{(KI)}_{n,m;\ell}(v)  - \text{p.v.} \int_{\Lambda^2}^{\Lambda^2_c} ds \, w_j^{(K)}(s) k_{j,\ell}^{(KI)}(s,v) 
\ee
Here repeated indices are summed over, meaning that $(Kj)$ and $(n,m)$ sums are cut by $J_c$ and $(n_c,m_c)$, and $(I\ell)$ sum goes up to $\infty$. 
The functional $d_\pm$ is to be minimized over the set of dual variables $\mathcal{D}$. However, to keep under control the infinite $\ell$ sum, one needs to make sure the squared expression in the first term of \reef{dualwithlambda} must be suppressed as $\ell \to \infty$. One way to achieve this is to minimize analytically for $\lambda_{(I\ell)}$, which gives 
\be
 2\lambda_{(I\ell)}/n_\ell = \sqrt{ \mu_{(I\ell)}(v)^2 + W_{(I\ell)}(v)^2 }\geq 0.
\ee
Plugging this expression in \eqref{dualwithlambda} produces
\bea
d_\pm(\{\nu,w\}) = \int^\infty_{\Lambda^2}  \frac{dv}{\rho(v)} &\sum^\infty_{(I\ell)} n_\ell 
\left[ \, 
	\mu_{(I\ell)}(v) +
	\sqrt{ \mu_{(I\ell)}(v)^2 + W_{(I\ell)}(v)^2 }
\, \right]
\label{duallag_on}
\eea
where we have defined
\be
W_{(I\ell)}(v) \equiv
\begin{cases} 
      w_{(I\ell)}(v)/n_\ell & \text{for } \ell \leq J_c \text{ and } v \leq \Lambda^2_c\\
      0 & \text{otherwise}
\end{cases}
\ee
and dropped the $\pm$ label on $d$ since its effect only enters into $c_H$ normalization condition, but not in the objective.
We stress that the functional $d_\pm$ must be evaluated on the solution of the two normalization conditions \eqref{clchcon}.

Now, a final remark is in order: note that in the region where $W_{(I\ell)}$ has no support, the integrand in the dual objective reduces to
$$ \mu_{(I\ell)}(v) \cdot \Theta \left[ \, \mu_{(I\ell)}(v) \, \right] $$
where $\Theta$ is the Heaviside Theta function. This means that higher spins won't contribute to the objective, as long as $\mu_{(I\ell)} \leq 0$.

\section{Numerical implementation}
\label{appni}

The (dual) problem presented before is a mathematically well-defined non-linear optimization problem, which can be studied by the favorite methods of the reader. For our purposes, we numerically looked for the minimum of \reef{duallag_on} and we chose to work with a linear problem solver named \texttt{SDPB}, mostly used in the conformal bootstrap literature \cite{Simmons-Duffin:2015qma,Landry:2019qug}. The reason of choice for us was its ability to achieve a reasonable degree of precision in the optimization objective, and the possibility to parallelize the computations on a cluster.

To put the problem on a computer, we need to take several steps to transform it into a suitable form. We start with turning the non-linear problem into a linear one, by the relaxation method.
 
\subsection{Relaxation}
Let us write a new objective function, by extending the set of dual variables to $$\mathcal{D}^\text{rel} = \mathcal{D} \cup \{\mathcal{X}^{\text{IR}}_{(I\ell)}(v), \mathcal{X}^{\text{UV}}_{(I\ell)}(v)\}$$
Then, the \emph{relaxed} dual objective, in which the new set of dual variables enter linearly:
\be
d^\text{rel}(\mathcal{D}^\text{rel}) = 
\int^{\Lambda_c^2}_{\Lambda^2} \frac{dv}{\rho^2(v)} \sum^{\infty}_{(I\ell)} n_\ell \, \mathcal{X}^{\text{IR}}_{(I\ell)}(v)
+
\int^{\Lambda_c^2}_{\Lambda^2} \frac{dv}{\rho^2(v)} \sum^{\infty}_{(I\ell)} n_\ell \, \mathcal{X}^{\text{UV}}_{(I\ell)}(v),
\label{duallag_relaxed_on}
\ee
subject to semi-positive conditions on the following $2\times2$ matrices
\bea
\label{const_relaxation_on}
\begin{pmatrix} \mathcal{X}^{\text{IR}}_{(I\ell)}(v) & W_{(I\ell)}(v) \\ W_{(I\ell)}(v) &  \mathcal{X}^{\text{IR}}_{(I\ell)}(v) - 2{\mu}_{(I\ell)}(v) \end{pmatrix} \succeq 0 \quad , \quad
\begin{pmatrix}  \mathcal{X}^{\text{UV}}_{(I\ell)}(v) & 0 \\ 0 &  \mathcal{X}^{\text{UV}}_{(I\ell)}(v) - 2{\mu}_{(I\ell)}(v) \end{pmatrix} \succeq 0
\eea
When these matrix constraints are saturated, r.h.s. of \reef{duallag_relaxed_on} reduces to r.h.s. of \reef{duallag_on}, and both dual objectives become equal $d^\text{rel}=d$. 

In this section, we assume the lower boundary of integration in \eqref{duallag_relaxed_on} to be the generic value $\Lambda^2$, instead of the normal threshold $4m^2$. This simple generalization will allow us the describe the setup discussed in Section \ref{eftb}, where we impose that $\Im M=0$ for $s<\Lambda^2$.

Notice that the positivity of the determinants imply $\mathcal{X}
^\text{IR}_\ell \geq \mu_\ell + \sqrt{\mu^2_\ell + (w_\ell/n_\ell)^2}$ and $\mathcal{X}^\text{UV}_\ell \geq 2 \mu_\ell \, \Theta[\mu_\ell]$. As a result, determinants measure how far away we are from saturating the true inequality $d^\text{rel} \geq d$. We will call the positive difference $d^\text{rel} - d$ measured by the determinant as \emph{the relaxation gap}.

The relaxation gap adds an additional layer of difficulty on the way of achieving the optimal solution to the dual problem.
However, it does not compromise the rigor of our approach, since its sole impact is to increase the objective, which still qualifies as a valid dual bound. To reduce the gap, we should increase the number of degrees of freedom in $\mathcal{X}^\text{IR}$, and $\mathcal{X}^\text{UV}$ variables as much as we can. 

\subsection{Spin cut-off}
\label{spincutoff}
We truncate the spin sum in \reef{duallag_on} to deal with only a finite number of spins. 
Remember that for $\ell > J_c$ the integrand for each spin becomes
$$ \mu_{(I\ell)} \cdot \Theta\left[ \mu_{(I\ell)} \right] $$
where $\Theta$ is the Heaviside-Theta function. This is interesting, because if we can show that there exists an $L_c$ such that $\mu_{(I\ell)}(v)<0$ for all $v \in [4,\infty]$ and $\ell > L_c > J_c$, the infinite tail would not contribute to the spin sum and we can safely truncate the sum at $L_c$! 

It turns out that there is such a corner in dual variables space, providing a feasible solution for our objective. The way we enforce the conditions are two-fold: (i) we study large-$\ell$ expansion of each term in $\mu_{(I\ell)}$, and make sure that it stays negative as $\ell \to \infty$.
(ii) for the intermediate spins, we impose by hand the negativity conditions. 
\begin{alignat}{2}
\text{(i)}:  \mu_{(I\ell)}(v) &\leq 0 \quad \text{for} \quad & L_c  &\leq \ell \label{const_infspin_on} \\
\text{(ii)}: \mu_{(I\ell)}(v) &\leq 0 \quad \text{for} \quad & J_c &\leq \ell \leq L_c \label{const_largespin_on}
\end{alignat}
By analyzing the asymptotics of (i) at large $\ell$ we find out that it is equivalent to the conditions
\be
\label{outer_const_on}
\renewcommand{\arraystretch}{1.25}
\begin{aligned}
 (-1)^{n_c+m_c} \, \cdot& \, y^{(sing)}<0 \\
 (-1)^{n_c+m_c} \, \cdot& \, y^{(sym)}<0 \\
 (-1)^{n_c+m_c+1} \, \cdot& \, y^{(anti)}<0 \\
\end{aligned} \quad \text{where} \quad 
\vec{y} \equiv
\left(
\begin{array}{ccc}
 n+1 & 1 & 1 \\
 \frac{2(n^2+n-2)}{n} & \frac{2(3 n-2)}{n} & \frac{2(-n-2)}{n} \\
 n-1 & -1 & 3 \\
\end{array}
\right) \cdot \vec{\nu}
\ee
and $\vec{v}\equiv(\nu^{(sing)}_{n_c,m_c} , \nu^{(sym)}_{n_c,m_c}, \nu^{(anti)}_{n_c,m_c})^T$, together with
\be
\label{inner_const_on}
\begin{aligned}
&\sideset{}{_{j=0}^{J_c} }\sum w^{(sym)}_j(\Lambda_c^2) P_j(0) > 0 \quad \text{with} \quad \Lambda_c^2 \in [v_c,4\Lambda^2{-}4] \\
&\sideset{}{_{j=0}^{J_c} }\sum \left( w^{(sing)}_j(\Lambda_c^2) - \frac{w^{(sym)}_j(\Lambda_c^2)}{n-1} \right) P_j(0) = 0
\end{aligned}
\ee
See next subsection for a detailed account on how to derive them.

We impose (i) through \reef{outer_const_on} and \reef{inner_const_on} and (ii) with a fixed $L_c$ on the relaxed problem, and we increase $L_c$ until a convergence in $d^\text{rel}$ is obtained. 
\subsubsection{Large $\ell$ asymptotics in more detail}
\label{largespin_on}
Large $\ell$ behavior of $\mu^{(I)}_{\ell}(v)$ is determined by Legendre polynomials in the definition of the kernels \eqref{eq:kernelsOn}. Their argument is either $x(v,t) \equiv 1+2t/(v-4)$ or $x(v,4/3)$. Remember that Legendre polynomials will grow exponentially whenever their argument exceed $\pm 1$. It turns out that there is a critical value 
\be
v_c = 2-4/3+\Lambda_c^2/2
\ee
along the integration range of $v$, such that 
\bea
|x(v,4/3)| \, &\geq |x(v,t)| & &\text{for} \quad v_c \leq v \\
|x(v,t)|   \, &\geq |x(v,4/3)| & &\text{for} \quad \Lambda^2  \leq v \leq v_c
\eea
We will call the two regions as outer and inner region respectively. \\
\textbf{Outer region $v_c < v$}. It is easy to see that at large $\ell$
\be
\overline{w}^{(I)}_\ell(v) \sim P_\ell \left( x(v,4/3) \right)
\ee
We have checked the crossing term for all $\textsf{n}+\textsf{m} \leq 7$, and the leading contributions go like
\be
F^{(KI)}_{\textsf{n,m};\ell}(v) \sim (-\ell)^{(\textsf{n}+\textsf{m}-2)} P_{\ell+\textsf{n}+\textsf{m}-2} \left( x(v,4/3) \right)
\ee
with a positive overall $\ell$-independent factor that we omitted.
Crossing symmetry contribution clearly wins over the other terms. We choose the set of crossing constraints such that 
$n_c+m_c$ gives uniquely the highest sum, and \reef{outer_const_on} will suffice to enforce $\mu^{(I)}_\ell<0$. \\
\textbf{Inner region $4 \leq v \leq v_c$}. 
The only dominant contribution in this case is
\bea
\overline{w}^{(I)}_\ell(v) &\sim P_\ell \left(x(v,t)\right)
\eea
$P_\ell(x)$ grows exponentially in $x<-1$, therefore $s$- and $z$-integrations in $\overline{w}^{(I)}_\ell(v)$ are dominated by the minimum of $x(v,t(s,z))$ which occurs at the end points $(s=\Lambda_c^2,z=0)$. Approximating the result using the saddle point method around the minimum gives
\be
\overline{w}^{(I)}_\ell(v) \approx \frac{1}{\ell^2} \frac{(\Lambda_c^2-v)^2}{\Lambda_c^2-4} \, \sum^{\textsf{J}_\text{max}}_{j,K} w^{(K)}_j(\Lambda_c^2) P_j(0) P_\ell \left( \frac{v-\Lambda_c^2}{v-4} \right) \mathbb{K}^{(KI)}(v;\Lambda_c^2,2{-}\Lambda_c^2/2)
\ee
Notice that $P_j(z\to0) \approx O(1)$ for $j$ even and $O(z)$ for $j$ odd. As a consequence, saddle point contributions from $w^{(anti)}_j$ are $O(1/\ell)$ suppressed with respect to symmetric and singlet channels. Therefore inner region constraints will be only on the dual variables $w^{(sing)}_j$ and $w^{(sym)}_j$.

Note further that $P_\ell\left( \tfrac{v-\Lambda_c^2}{v-4} \right) > \pm 1$ for all $v$ and even/odd $\ell$. To enforce $\mu^{(I)}_\ell<0$, we need to combine $\overline{w}^{(sing,sym)}_\ell > 0$ and $\overline{w}^{(anti)}_\ell < 0$. These three conditions put together eventually imply the ones in \reef{inner_const_on}.
\subsection{Dual Ansatzes}
\label{ansatzes}
Next, we describe how to write an ansatz for the dual variables in terms of a finite basis of functions. We send the interval $x \in [-1,1]$ into $v^\text{IR} \in [\Lambda^2,\Lambda^2_c]$ and $v^\text{UV} \in [\Lambda^2_c,\infty)$ with the following maps.
\bea
v^\text{IR}(x) 
= \frac{1}{2}(\Lambda^2_c+\Lambda^2) +\frac{x}{2} (\Lambda^2_c-\Lambda^2) \quad , \quad
v^\text{UV}(x) = 
\frac{\pi}{3} \Lambda^2_c \tan \left(\frac{\pi}{4} (x+1)\right) \sec ^2\left(\frac{\pi}{4} (x+1)\right) \, .
\eea
Then we parametrize the dual variables $\{\mathcal{X}^{\text{IR}}_\ell, \mathcal{X}^{\text{UV}}_\ell, w_\ell \}$ in terms of Chebyshev polynomials $T_n$
\bea
w^{(I)}_\ell(x) = \sum^{N_w}_{n=0} a^{(I)}_{\ell,n} \, T_n(x) \quad , \quad \mathcal{X}^{(I),\text{IR}}_\ell(x) = \sum^{N_\text{IR}}_{n=0} c^{(I)}_{\ell,n} \, T_n(x) \quad , \quad \mathcal{X}^{(I),\text{UV}}_\ell(x) = \sum^{N_\text{UV}}_{n=0} d^{(I)}_{\ell,n} \, T_n(x)
\label{dual_ansatzes_on}
\eea
such that $dx \, \mathcal{X}_\ell(x) = dv \, \mathcal{X}_\ell(v)$ and they give the same result under the integral sign. 
In our runs, we will often take $N_\text{max}=N_w=N_\text{IR}=N_\text{UV}$ and we sample above functions of $x$ on a Chebyshev grid with 199 points on $[-1,1]$ for both IR and UV sections.

Notice that $dv^\text{UV}/dx$ has a zero at $x=-1$ which can cause us problems, since
$$
\lim_{v\to\Lambda_c^2} \mathcal{X}^{\text{UV}}_\ell(v) = \lim_{x\to-1} \mathcal{X}^{\text{UV}}_\ell(x) \left[ \frac{dv^\text{UV}}{dx} \right]^{-1}
$$
will behave singular, unless $\mathcal{X}^{\text{UV}}_\ell(x) \xrightarrow{x\to-1} O(x+1)$. To make sure it is regular, we will require
\be
\mathcal{X}^{(I),\text{UV}}_\ell(-1) = \sum^{N_\text{UV}}_{n=0} d^{(I)}_{\ell,n} \, T_n(-1) = 0 \, . \label{continuity_on}
\ee
All in all, the dual problem to be solved numerically is the following: 
\be
\label{radial_opt_dual_relaxed_ans_on}
\begin{aligned}
\min \quad d^\text{rel}(\mathcal{D}^\text{rel}) \qquad &\text{over} \quad \{ \nu^{(I)}_\textsf{n,m}, a^{(I)}_{\ell,n},c^{(I)}_{\ell,n},d^{(I)}_{\ell,n} \} \\
&\text{subject to} \quad \{ \, \reef{clchcon}, \, \reef{const_relaxation_on} \, , \, \reef{const_largespin_on}, \, \reef{outer_const_on}, \, \reef{inner_const_on}, \, \reef{continuity_on} \, \}
\end{aligned}
\ee

\subsection{The space of $O(n)$ theories}
\begin{figure}[h] \centering
\includegraphics[scale=.45]{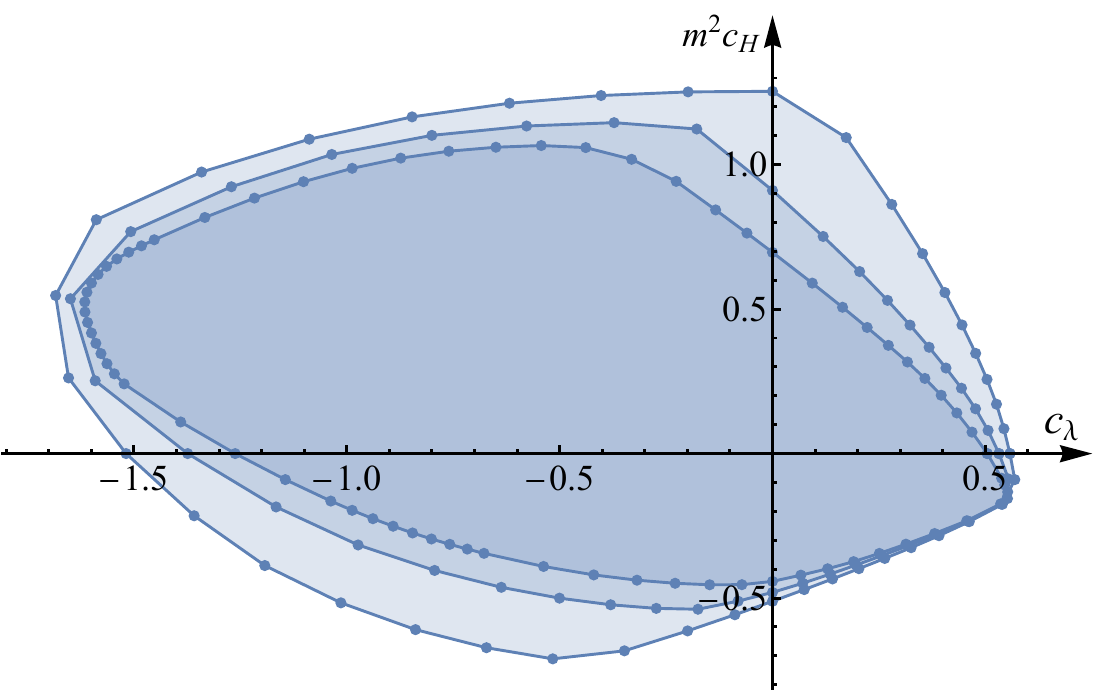} 
\caption{
A family of $O(n)$ dual exclusion plots, for $n=2,3,4$ respectively from outer to innermost boundary. Markers along the boundaries indicate the angular grid chosen in the radial optimization problem described in section~\ref{islando4}.
}
\label{fig:othern}
\end{figure}

We have left the parameter $n>1$ to be a generic integer so far. In order to explore the space of nonperturbative islands at various $n$, we solved the radial optimization problem given in section~\ref{islando4} for a couple of values $n=2,3,4$. The resulting islands are shown in fig.~\ref{fig:othern}.

Numerical bounds we obtain in fig.~\ref{fig:dual_primal_oyster}, fig.~\ref{fig:interpolation} and fig.~\ref{fig:othern} depend on the cutoff parameters $\{J_c, L_c, (n_c,m_c), N_\text{max}\}$. 
To obtain well converged bounds, we fixed them to the following values in all of our runs:
$$J_c = 5 \quad , \quad L_c= 25 \quad , \quad (n_c,m_c) = (1,6) \quad , \quad N_\text{max}=8 \, .$$

\section{Perturbative computations}

\subsection{Perturbative $\lambda |\vec \phi|^4$ low energy constants }
\label{lecs}

Consider the interaction Lagrangian $\mathcal{L}_\text{int}=- \lambda |\vec \phi|^4/8$. The amplitude at tree-level is given by $M=-\lambda$. 
Here, we are interested in computing the leading order contribution to the low energy constants coming from the one-loop interaction.
The bare coupling $\lambda$ diverges at one-loop, so we just redefine the coupling using the physical amplitude $c_\lambda=M(s=t=u=4/3)=-\lambda_R$. The imaginary part at one-loop is simply given by
\be
\Im \vec{M}(s \geq 4m^2,t) =\frac{1}{2} \frac{\lambda^2}{16\pi} \sqrt{\frac{s-4m^2}{s}} \, \big((n+2)^2, 4, 0 \big).
\ee
Plugging this expression into the dispersive representations \eqref{chsr}, \eqref{c2sumrule}, and \eqref{c2primesumrule}, we obtain 
\bea
\label{oneloop-coeffs}
c_H &= \frac{c_\lambda^2}{256 \pi^2 } \, 3 (n+2) \left(3 \sqrt{2} \tan ^{-1}\left(\frac{1}{\sqrt{2}}\right)-2\right), \\
c_2 &= \frac{c_\lambda^2}{4096 \pi^2 } \, 9 \left(14-15 \sqrt{2} \tan ^{-1}\left(\frac{1}{\sqrt{2}}\right)\right), \\
2c_2+c'_2 &= \frac{c_\lambda^2}{8192 \pi^2 } \, 9 (n+8) \left(14-15 \sqrt{2} \tan ^{-1}\left(\frac{1}{\sqrt{2}}\right)\right).
\eea

\subsection{Further comments on  \reef{tuning}}
\label{apptun}

We will now present a simple  perturbative calculation to gain additional insights into the range of validity of \reef{tuning}.
Consider the Lagrangian ${\cal L}= ( \partial\vec \phi)^2/2- m^2\vec \phi^2/2 - \lambda \vec \phi^4/8 -M^2 \Phi^2/2 - g \vec \phi^2 \Phi/2 +O(\Phi^3)$, with   $M^2\gg m^2$. The imaginary part of the  two-to-two scattering amplitude, at lowest non-trivial order,  is given by
\be
\Im \vec{M}(s \geq 4m^2,t) = \  \diagIRone  \ + \ \diagUVtwo =
\frac{1}{2} \frac{\lambda^2}{16\pi} \sqrt{\frac{s-4m^2}{s}} \, \big( (n+2)^2, 4, 0 \big)
+ g^2 \pi \delta(s-M^2) \, \big( n, 0, 0 \big)
\label{pertuvamp}
\ee
which has non-vanishing partial wave projection to spin-zero only, in both singlet and symmetric isospin channels.
Then, the only integrals to evaluate in the sum rule \reef{chsr} are for $J=1,2$ and $\ell=0$. The corresponding kernels are 
\be
\label{chsrkernels}
\kappa^{(3,1)}_{1,0}(s,v) = -\frac{1}{3}\kappa^{(3,2)}_{1,0}(s,v) =
\frac{-12 (s-4) \left(v-\frac{4}{3}\right)^2+6 \left(v-\frac{4}{3}\right)^2 (s+2 v-4) \log \left(\frac{s+v-4}{v}\right)-(s-4)^3}{192 \pi ^2 (s-4)^2 \left(v-\frac{4}{3}\right)^2}
\ee
Let us choose $\Lambda^2 = M^2$ and study the following IR/UV ratio
\be
r[s] \equiv \frac{c_H^\text{IR}(s)}{c_H^\text{UV}(s)} \quad \text{with} \quad
c_H^\text{IR}(s) \equiv \int_{4}^{\Lambda^2} dv \, k^{(3,J)}_{1,\ell}(s,v) \text{Im}f^{(J)}_\ell(v) \quad , \quad
c_H^\text{UV}(s) \equiv \int_{\Lambda^2}^\infty dv \, k^{(3,J)}_{1,\ell}(s,v) \text{Im}f^{(J)}_\ell(v) \, ,
\ee
to be evaluated on the above amplitude. We will fix the subtraction point to two values: $s=4m^2$ and $s=M^2$.

\begin{figure}[t]
\centering
\includegraphics[scale=.45]{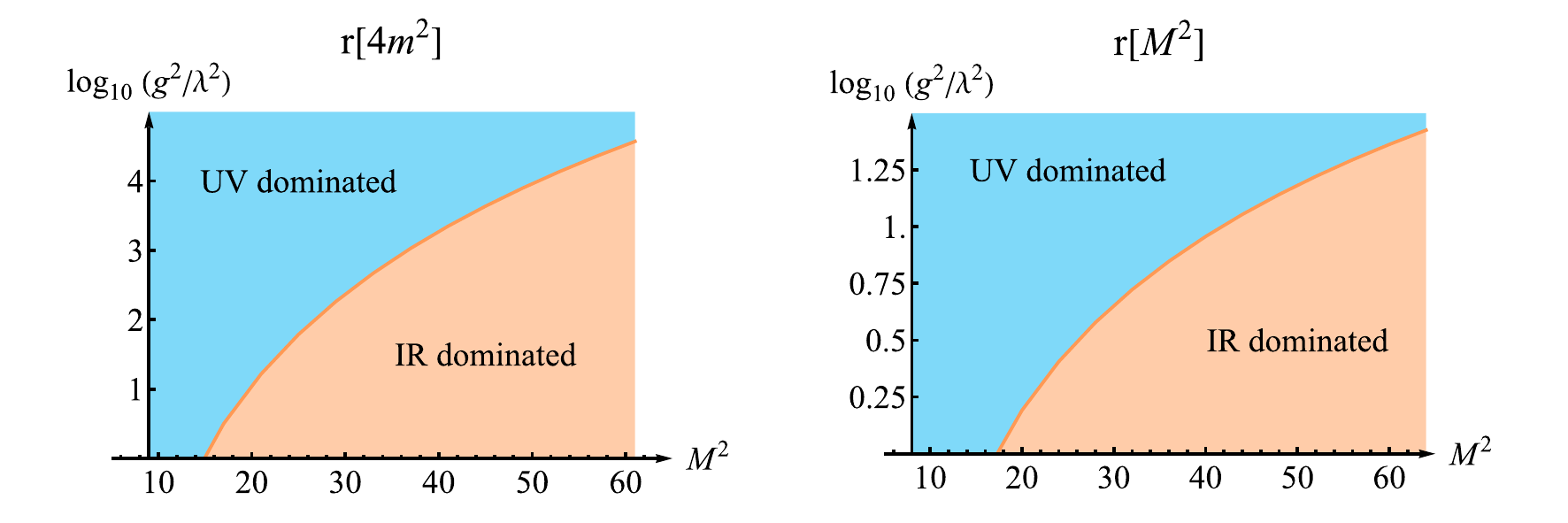} 
\caption{
IR/UV domination regimes of the amplitude \reef{pertuvamp}. Blue is when $r<1$ and orange is when $r>1$.
}
\label{fig:pertuv}
\end{figure}

$r[s]$ then depends on two free parameters of the amplitude \reef{pertuvamp}: The ratio of couplings $g^2/\lambda^2$ and the mass of the heavy particle $M^2$. Defining $r[s]=1$ as the transition point between IR/UV domination regimes, we find out the domination regions in the parameter space as shown in fig.~\ref{fig:pertuv}.

\section{Rigorous bounds assuming an IR model}
\label{appdir}

In this Appendix we solve the problem discussed in~\ref{IRmodel}. We assume that
\be
\Im f_\ell^{(I)}(s)=g_\ell^{(I)}(s), \quad s\leq \Lambda^2.
\ee
The phenomenological input enters into the choice of $\Lambda$ and $g_\ell(s)$.
With this condition the Roy equations are given by
\be
\Re f_\ell^{(I)}(s)=c_\ell^{(I)}(s)+\frac{1}{\pi}\int_{4m^2}^{\Lambda^2}\sum_{j,K} g_j^{(K)}(v)k_{j\ell}^{(K I)}(v,s)dv+\frac{1}{\pi}\int_{\Lambda^2}^{\infty}\sum_{j,K} \Im f_j^{(K)}(v)k_{j\ell}^{(K I)}(v,s)dv.
\ee
We want to solve the problem of maximizing $\alpha c_H$ where $\alpha=\pm1$ within our model.
We write the Lagrangian (we omit the unitarity constraints for simplicity)
\be
\mathcal{L}=\alpha c_H+\sum_{\ell I}^L \int_{\Lambda^2}^{\mu^2} \Re f_\ell^{(I)}(s)w_\ell^{(I)}(s)ds+\sum_{\ell I}^\infty \int_{\Lambda^2}^{\infty} \Im f_\ell^{(I)}(s) \bar w_\ell^{(I)}(s)ds+\sum_{\ell I}^\infty \int_{4m^2}^{\Lambda^2} g_\ell^{(I)}(s) \bar w_\ell^{(I)}(s)ds+\sum_{\ell I}^L \int_{\Lambda^2}^{\mu^2} c_\ell^{(I)}(s)w_\ell^{(I)}(s)ds,
\ee
where 
\be
\bar w_j^{(K)}(v)=-\frac{1}{\pi} \sum_{\ell I}^L \int_{\Lambda^2}^{\mu^2}w_\ell^{(I)}(s) k_{j\ell}^{(K I)}(v,s) ds.
\ee
Maximizing w.r.t the primal variables $c_\lambda, c_H$ we obtain the dual constraints
\be
\frac{\partial}{\partial c_H}\left(\alpha c_H+\sum_{\ell I}^L \int_{\Lambda^2}^{\mu^2} c_\ell^{(I)}(s)w_\ell^{(I)}(s)ds\right)=0, \quad \frac{\partial}{\partial c_\lambda}\left(\alpha c_H+\sum_{\ell I}^L \int_{\Lambda^2}^{\mu^2} c_\ell^{(I)}(s)w_\ell^{(I)}(s)ds\right)=0,
\ee
while first maximizing w.r.t the physical partial waves, and then minimizing w.r.t. the unitarity constraints we obtain the inequality
\be
c_H\leq D=D^{UV}+D^{IR}[g^{(I)}_\ell],
\ee
where
\be
D^{UV}=\sum_{\ell I}^L \int_{\Lambda^2}^{\mu^2} \frac{\bar w_\ell^{(I)}(s)+\sqrt{(\bar w_\ell^{(I)}(s))^2+ (w_\ell^{(I)}(s))^2}}{\rho^2(s)}ds,
\ee
and
\be
D^{IR}[g^{(I)}_\ell]=\sum_{\ell I}^\infty \int_{4m^2}^{\Lambda^2} g_\ell^{(I)}(s) \bar w_\ell^{(I)}(s)ds,
\ee
provided that $\bar w_\ell^{(I)}(s)\leq 0$ whenever it does not appear in $D^{UV}$ or $D^{IR}$.

For each given IR model we can solve a dual problem. Let's discuss now some limiting situations.
Imagine $g^{(I)}_\ell=0$, which gives 
\be
c_H\leq D^{UV}.
\ee
This approximation can be well justified in two scenarios.
One is realized when we have weakly coupled UV complete models such as in gauge theories with large $N_c$. In this case we do expect our bound to be extremely loose. The second scenario is realized when, due to non-perturbative effects, there is a cancellation among terms in the low energy expansion and we can neglect the imaginary parts way beyond the radius of convergence of the EFT. This scenario sometimes is realized on the boundary of the allowed region determined by non-perturbative Bootstrap studies~\cite{EliasMiro:2022xaa}. 

Suppose now that $D^{IR}[g^{(I)}_\ell]\leq 0$. Then, we obtain the chain of inequalities
\be
c_H\leq D^{UV}+D^{IR}[g^{(I)}_\ell]\leq D^{UV}.
\ee
In it could be possible to impose this condition, the bound should hold for any IR model, and, therefore contain the $O(4)$ island.
In general, however, it is hard to satisfy the inequality $D^{IR}[g^{(I)}_\ell]\leq 0$, and, a priori, we cannot rigorously use the $D^{UV}$ functional alone to bound the Wilson coefficients in presence of an IR imaginary part.

However, it is possible to obtain a bound on $c_H$ by solving first the truncated optimization problem
\be
d^{UV}=\min_w D^{UV},
\label{UV_universal}
\ee
which is attained for some critical $w_c$, then construct the bound
\be
c_H\leq D^{IR}[g^{(I)}_\ell]+d^{UV}
\ee
by plugging $w_c$ in $D^{IR}[g^{(I)}_\ell]$.

So, solving the simple universal problem \ref{UV_universal} -- see also \ref{eftb},  has a conceptual value since it can be used to generate rigorous bounds for any choice of the $g_\ell^{(I)}$.
If we commit from the beginning with some IR model, then we can fully optimize $D^{UV}+D^{IR}[g^{(I)}_\ell]$, and obtain even stronger bounds.

\section{Phase-shifts}
\label{psapp}

We can reconstruct the optimal dual S-matrices from the partial waves $f^{(I)}_\ell(v)$ on the support where we impose the Roy equations.
 $w^{(I)}_\ell$ exist, because then we can use the fixed-$t$ dispersion relation to reconstruct $\text{Re}f^{(I)}_\ell(v)$, as can be seen in \reef{dualpwaves}. 
 Then the S-matrix on a single partial wave channel is given by
\bea
S^{(I)}_\ell(s) = 1 + i \sqrt{\frac{s-4}{s}} f^{(I)}_\ell(s) 
\eea
which is a pure complex phase evaluated on the solutions \reef{dualpwaves}. This allows us then to plot \emph{the phase shift} of the scattered wave as a function of $s$
\be
 \delta^{(I)}_\ell(s) = \frac{1}{2i} \log S^{(I)}_\ell(s)
\ee
Below we give sample dual phase shifts along the four distinct sections of the $O(4)$ nonperturbative island. Note that a threshold singularity puts $\delta^{(I)}_\ell(0) = \pi/2$ which would otherwise be zero.
\begin{figure}[h] \centering
	\includegraphics[scale=.5]{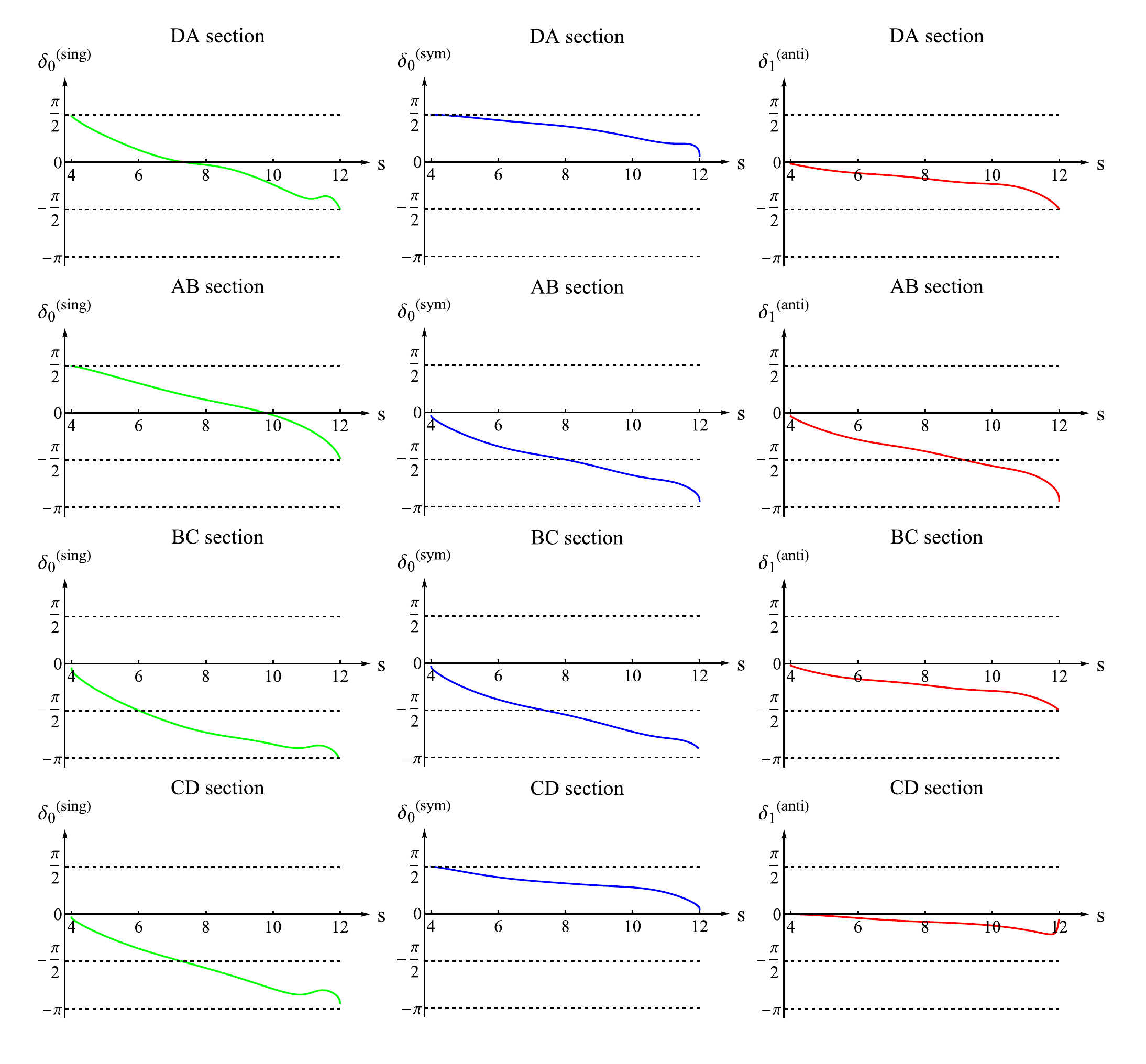} 
	\caption{Sample phase shifts from each of four sections along the boundary of Figure \ref{fig:dual_primal_oyster}.}
\end{figure}

\twocolumngrid

\end{document}